\documentclass[11pt,a4paper]{article}

\usepackage[bookmarks=true,bookmarksnumbered=true,setpagesize=false]{hyperref}

\usepackage[utf8]{inputenc}
\usepackage{amsmath,amsfonts,amssymb,amsthm,color,enumerate}
\usepackage{graphicx}
\usepackage{url}

\usepackage{color}
\usepackage[dvipsnames]{xcolor}
\usepackage{braket}

\usepackage{comment}
\usepackage{a4wide}

\renewcommand{\arraystretch}{2}

\newcommand{\tx}{\text}
\newcommand{\ov}{\over}

\newcommand{\MP}{\ensuremath{M_\text{P}}}
\newcommand{\MR}{\ensuremath{M_\text{R}}}
\newcommand{\eff}{\text{eff}}

\newcommand{\red}[1]{#1}
\newcommand{\blue}[1]{#1}
\newcommand{\magenta}[1]{#1}

\newcommand{\paren}[1]{\left(#1\right)}
\newcommand{\fn}[1]{\!\left(#1\right)}

\newcommand{\df}{\text{d}}

\newcommand{\al}[1]{\begin{align}#1\end{align}}

\newcommand{\nn}{\nonumber\\}

\newcommand{\eV}{\,\text{eV}}
\newcommand{\GeV}{\,\text{GeV}}
\newcommand{\TeV}{\,\text{TeV}}

\newcommand{\bi}{\begin{itemize}}
\newcommand{\ei}{\end{itemize}}

\newcommand{\ef}[1]{#1}

\begin{document}
\title{
\vspace{-2cm}
\vbox{
\hfill \hbox{\normalsize OU-HET/948}\\
\vspace{2cm}
}
{\bf 
\magenta{Cosmological implications of
Standard Model criticality and Higgs inflation}
}
}

\author{
	Yuta Hamada,\thanks{\tt
		yhamada@wisc.edu}
		$^{1,2}$\
	Hikaru Kawai,\thanks{\tt
		hkawai@gauge.scphys.kyoto-u.ac.jp
		}
		$^3$\
	Yukari Nakanishi,\thanks{\tt
		nakanishi@het.phys.sci.osaka-u.ac.jp
		}
		$^4$\
	and
	Kin-ya Oda\thanks{\tt
		odakin@phys.sci.osaka-u.ac.jp
		}
		$^4$\\
$^1$\it\normalsize
Department of Physics, University of Wisconsin, Madison, WI 53706, USA\\
$^2$\it\normalsize
KEK Theory Center, IPNS, KEK, Tsukuba, Ibaraki 305-0801, Japan\\
$^3$\it\normalsize
Department of Physics, Kyoto University, Kyoto 606-8502, Japan\\
$^4$\it\normalsize
Department of Physics, Osaka University, Osaka 560-0043, Japan
	}
\date{\today}
\maketitle

\begin{abstract}\noindent
\magenta{The observed Higgs mass indicates that the Standard Model can be valid up to near the Planck scale $M_\text{P}$. Within this framework, it is important to examine how little modification is necessary to fit the recent experimental results in particle physics and cosmology. As a minimal extension, we consider the possibility that the Higgs field plays the role of inflaton and that the dark matter is the Higgs-portal scalar field. 
We assume that the extended Standard Model is valid up to the string scale $10^{17}\,\text{GeV}$.
(This translates to the assumption that all the non-minimal couplings are not particularly large, $\xi\lesssim 10^2$, as in the critical Higgs inflation, since $M_\text{P}/\sqrt{10^2}\sim 10^{17}\,\text{GeV}$.)
We find a correlated theoretical bound on the tensor-to-scalar ratio $r$ and the dark matter mass $m_\text{DM}$.
As a result,}
the Planck bound $r<0.09$ implies that the dark-matter mass must be smaller than 1.1\,TeV,
while the PandaX-II bound on the dark-matter mass $m_\text{DM}\magenta{>0.7\pm0.2\,\text{TeV}}$ leads to $r\gtrsim \magenta{2}\times10^{-3}$.
Both are within the range of near-future detection.
When we include the right-handed neutrinos of mass $M_\text{R}\sim 10^{14}$\,GeV, the allowed region becomes wider, but we still predict $r\gtrsim 10^{-3}$ in the most of the parameter space. The most conservative bound becomes $r>10^{-5}$ if we allow three-parameter tuning of $m_\text{DM}$, $M_\text{R}$, and the top-quark mass.

\end{abstract}


\newpage

\tableofcontents

\newpage

\section{Introduction}\label{intro}

\begin{figure}[t]
  \begin{center}
   \includegraphics[height=8em]{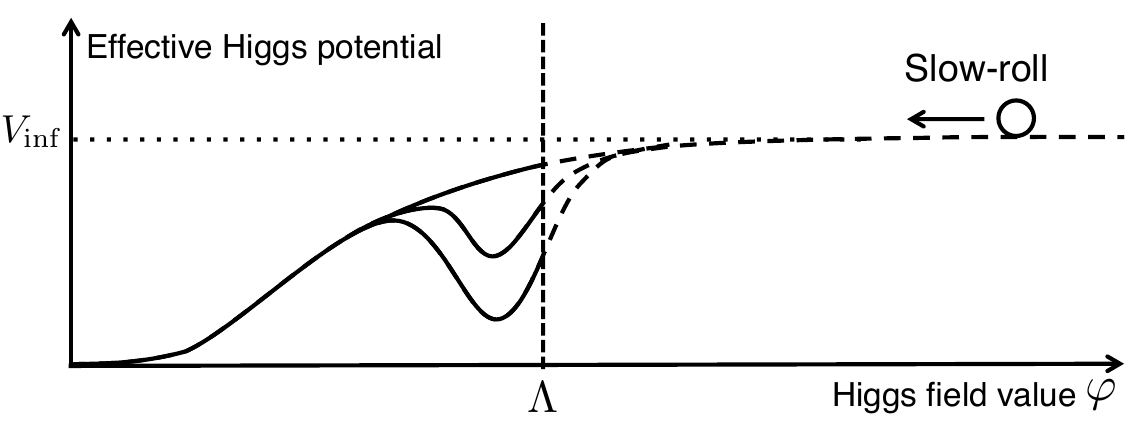}
  \end{center}
  \caption{Schematic figure for the Higgs field as an inflaton}
  \label{fig:inflation}
\end{figure}

The Higgs field is the only elementary scalar whose existence is experimentally confirmed.\ef{\footnote{\ef{
There always remains a possibility that the Higgs field turns out to be a composite field in an unexplored energy region; see e.g.\ Refs.~\cite{Sanz:2017tco,Witzel:2019jbe} for recent reviews.
}}}
\magenta{The observed Higgs mass indicates that the Standard Model (SM) can be valid up to near the Planck scale $\MP=1/\sqrt{8\pi G}\simeq 2.4\times10^{18}\GeV$, that is, all the couplings remain perturbative and the Higgs potential is stable; see e.g.\ Refs.~\cite{Holthausen:2011aa,Bezrukov:2012sa,Degrassi:2012ry,Alekhin:2012py,Masina:2012tz,Hamada:2012bp,Jegerlehner:2013cta,Buttazzo:2013uya,Branchina:2013jra,Kobakhidze:2014xda,Spencer-Smith:2014woa,Branchina:2014usa,Branchina:2014rva,Bednyakov:2015sca}.\ef{\footnote{
\ef{See Refs.~\cite{Casas:1999cd,EliasMiro:2011aa,Chen:2012faa,Rodejohann:2012px,Masina:2012tz} for the renormalization group analyses on the vacuum stability including the right-handed neutrinos.}
}}
From string-theory point of view, this fact suggests that string theory is directly connected to the SM at the string scale $\Lambda\sim 10^{17}\,\text{GeV}$.

Furthermore, the Higgs potential can be very small and flat around the Planck scale \magenta{by tuning the top quark mass within the experimental error}. 
This fact, the so-called criticality of the SM, suggests that something non-trivial is happening around the Planck scale and that the SM remains valid without much modification up to the scale.\footnote{\magenta{
This paradigm includes e.g.\ the multiple point criticality principle (MPP)~\cite{Froggatt:1995rt,Froggatt:2001pa,Nielsen:2012pu,Haba:2016gqx,Haba:2017quk}, the (classical) conformality~\cite{Meissner:2007xv,Foot:2007iy,Meissner:2007xv,Iso:2009ss,Iso:2009nw,Hur:2011sv,Iso:2012jn,Hashimoto:2014ela,Chankowski:2014fva,Kobakhidze:2014afa,Gorsky:2014una,Kubo:2014ova,Foot:2014ifa,Kawana:2015tka,Latosinski:2015pba,Haba:2015nwl,Haba:2015qbz,Kubo:2016kpb,Haba:2017wwn,Iso:2017uuu,Lewandowski:2017wov}, the asymptotic safety~\cite{Shaposhnikov:2009pv,Wetterich:2016uxm,Eichhorn:2017als}, the hidden duality and symmetry~\cite{Kawamura:2013kua,Kawamura:2013xwa}, and the maximum entropy principle~\cite{Kawai:2011qb,Kawai:2013wwa,Hamada:2014ofa,Kawana:2014vra,Hamada:2014xra}.}}
Within this paradigm, it is important to examine how little modification is necessary to fit the recent experimental results in particle physics and cosmology, especially the inflation and the dark matter.

%
The flatness of the Higgs potential suggests that the Higgs field can play the role of inflaton. 
Indeed, if we trust the SM even at the Planck scale, we can realize a phenomenologically viable inflation, namely the critical Higgs inflation, by introducing a non-minimal coupling of order 10--$10^2$~\cite{Hamada:2014iga,Bezrukov:2014bra,Hamada:2014wna}; see also Ref.~\cite{Cook:2014dga}.\footnote{\magenta{
In the original Higgs inflation model~\cite{Salopek:1988qh,Bezrukov:2007ep}, the flatness of the Higgs potential has not been assumed and a large non-minimal coupling $\xi\sim 10^{4\tx{--}5}$ has been required.}
}

In this paper, we \blue{\emph{do not assume any particular form}} of the Higgs potential at the Planck scale.\footnote{\magenta{
If one assumes e.g.\ that the Higgs field fits in the massless state in string theory, one may compute the potential at the trans-Planckian scale and discuss the possibility of its mixing with other directions such as the radius of the extra dimension at the high scales~\cite{Hamada:2013mya,Hamada:2015ria}.
}}
Instead, we study consequences of a general postulate that the Higgs field plays the role of inflaton above the string scale, assuming that the SM (extended with dark matter) is reliable below it; see Fig.~\ref{fig:inflation}.\footnote{\label{Higgs inflation footnote}
\magenta{This is consistent with the natural assumption that all the non-minimal couplings are not particularly large, namely, at most $10^2$, as in the critical Higgs inflation.
}}
}
After the end of the inflation, the slow-roll condition on the Higgs field is violated.
In order for the fields to roll down to the electroweak (EW) scale, the potential height must be smaller than the inflation energy $V_\tx{inf}$ in the whole region $\varphi\leq\Lambda$.
Note that even if there exists a local maximum with its height smaller than $V_\tx{inf}$, it does not interrupt the rolling down to the EW scale because the slow-roll condition is already violated.
As we do not specify the shape of the inflaton potential above $\Lambda$, we cannot predict precisely the cosmological parameters such as the spectral index $n_s$ and the tensor-to-scalar ratio $r$.
However, we may still put a lower bound on $V_\tx{inf}$ from the highest value of the Higgs potential in the region $\varphi<\Lambda$, which can be converted into the lower bound on $r$.

It is certain that there exists a dark matter (DM).
As one of the simplest realization, we employ the Higgs portal $Z_2$ scalar dark matter model~\cite{Silveira:1985rk,McDonald:1993ex,Burgess:2000yq,Davoudiasl:2004be,Patt:2006fw,Grzadkowski:2009mj,Drozd:2011aa,Haba:2013lga,Baek:2014jga},\footnote{
See Ref.~\cite{Ko:2014eia} for a model with extra dark Higgs field.
}
though our analysis itself is applicable for any other model that modifies the running of the Higgs quartic coupling.
We consider the generic region of the DM mass $m_\tx{DM}$ being larger than the Higgs mass.
Then its thermal abundance fixes the relation between the DM mass and the Higgs-DM coupling $\kappa$ to be $m_\tx{DM}\simeq\kappa\times3.2\TeV$,
and the spin-independent DM-nucleon elastic cross section is determined to be $\sigma_\tx{SI}\sim10^{-45}\,\tx{cm}^2$~\cite{Cline:2013gha,Hamada:2014xka}; see Fig.~\ref{fig:DM constraint} for more discussion.
The latest 1.6\,$\sigma$ bound from 
\magenta{the PandaX-II experiment~\cite{Cui:2017nnn} reads $m_\tx{DM}\gtrsim0.7\TeV$.}
When we do not include the right-handed neutrinos, we find that the current observational bounds $r<0.09$ and $m_\tx{DM}\gtrsim \magenta{0.7\TeV}$ lead to the theoretical bounds $m_\tx{DM}\lesssim 1.1\TeV$ and $r\gtrsim \magenta{2\times}10^{-3}$, respectively, which are well within the range of near future detection.

We have also studied the case with the right-handed neutrinos that account for the observed neutrino oscillations through the seesaw mechanism~\cite{Minkowski:1977sc,seesawOthers}.
We find that when their mass is in the range $\MR\lesssim 10^{13}\GeV$, the results are the same as in the case without them.
As we increase $\MR$, the bound becomes milder up to the scale $10^{14}\GeV$, and then becomes tighter up to $10^{15}\GeV$ at which the right-handed neutrino contribution makes the Higgs potential unstable.
Combining these results, we find the absolute theoretical bound $r>10^{-5}$.
If we restrict $m_\tx{DM}\gtrsim 1.3\TeV$, we obtain a stronger bound $r\gtrsim 10^{-3}$ for a reasonable top-quark mass range, as we will see in Fig.~\ref{fig:houraku2sigma}.

This paper is organized as follows. 
In Sec.~\ref{higgsinf}, we show our basic strategy how to put a lower bound on $r$ without the knowledge of higher scales.
In Sec.~\ref{darkmatter}, we review the Higgs-portal $Z_2$ scalar dark matter.
In Sec.~\ref{neutrinoless section}, we show our results without the effects from the heavy right-handed neutrinos.
In Sec.~\ref{neutrino section}, we show the results with the right-handed neutrinos for several representative values of their mass $\MR$.
In Sec.~\ref{sec:all}, we show the allowed region when we vary both $\MR$ and the top-quark mass $m_t$.
In Sec.~\ref{summary section}, we summarize and discuss our results.
In Appendix~\ref{calculation}, we show the renormalization group equations (RGEs) that we employ in the computation of the effective potential.
In Appendix~\ref{OHP}, we discuss that the shape of allowed region, shown in Secs.~\ref{neutrinoless section}--\ref{sec:all}, can be understood in terms of the difference of potential shape.

\section{Lower bound on tensor-to-scalar ratio}\label{higgsinf}
We present our basic strategy how to put a lower bound on $r$ \blue{\emph{without the knowledge of the physics at the higher Higgs-field value}} $\varphi>\Lambda$, extending the analysis in Ref.~\cite{Hamada:2013mya}.
In the slow-roll inflation, the observable $A_s$ and $r$ are written in terms of $\epsilon_V$ and $V_\tx{inf}$:
\al{
A_s
	&=	{1\over24\pi^2}{1\over\epsilon_V}{V_\tx{inf}\over\MP^4},	&
r	&=	16\epsilon_V,
}
where
\al{
\epsilon_V:={\MP^2\over2}\paren{V_{,\varphi}\over V}^2,
}
is the slow-roll parameter. Eliminating $\epsilon_V$, we obtain
\al{
r	&=	{2\over3\pi^2}{1\over A_s}{V_\tx{inf}\over\MP^4}.
		\label{v-and-r}
}
This gives a linear relation between $r$ and $V_\tx{inf}$ 
since $A_s$ is fixed by the CMB observation to be $A_s\simeq2.2\times10^{-9}$~\cite{Ade:2015lrj}.

During the inflation, the inflaton field value is larger than $\Lambda$; see Fig.~\ref{fig:inflation}.
After the end of inflation, the field continues to roll down the potential hill and becomes the low-energy Higgs field that we know in the SM.
In order not to prevent the rolling down to the EW scale, the maximum value of the effective potential in the region $\varphi\leq\Lambda$, which we call $V_{\varphi\leq\Lambda}^\tx{max}$, must be smaller than the energy density during the inflation $V_\tx{inf}$:
\al{
V_{\varphi\leq\Lambda}^\tx{max}< V_\tx{inf}.
\label{VmaxVinf}
}
\blue{We may rewrite Eq.~\eqref{VmaxVinf}, using Eq.~\eqref{v-and-r}, into the form:}
\al{
r> {2\over3\pi^2}{1\over A_s}{V_{\varphi\leq\Lambda}^\tx{max}\over\MP^4}.
\label{eq:rmin}
}
Thus, we obtain the lower bound on $r$ from $V_{\varphi\leq\Lambda}^\tx{max}$ only.
\blue{For later convenience, we name the combination that appears in the right-hand side of Eq.~\eqref{eq:rmin} $r_\tx{bound}$:
\al{
r_\tx{bound}:={2\over3\pi^2}{1\over A_s}{V_{\varphi\leq\Lambda}^\tx{max}\over\MP^4}.
	\label{definition of r_bound}
}
We emphasize that $V_{\varphi\leq\Lambda}^\tx{max}$ is obtained from the physics at low-energy scales, while $A_s$ and $r$ in Eqs.~\eqref{eq:rmin} and \eqref{definition of r_bound}, respectively, are the usual observable quantities that are defined at the horizon exit of the wavelength of the order of the CMB scale.}

\red{
$r_\text{bound}$ roughly scales as $\propto \Lambda^4$ so that one can rescale our results accordingly if needed. Our choice $\Lambda=10^{17}$\,GeV
is motivated by the following facts:
\begin{itemize}
\item This value is (slightly below) the critical point at which both the Higgs quartic coupling and its beta function becomes zero; see e.g.\ Ref.~\cite{Hamada:2012bp}.\footnote{\red{
If we instead used $\lambda_\text{eff}$, which is the Higgs effective potential divided by $\varphi^4$~\cite{Hamada:2014wna}, $\Lambda$ would be $\Lambda\sim10^{18}$\,GeV, in which case the lower bound on $r$ would become more stringent. We stay on the conservative side throughout this paper.
}}
\item This is the standard perturbative string scale; see any textbook of string theory or e.g.\ references from Ref.~\cite{Hamada:2015ria}.
\end{itemize}
In the Higgs inflation, as we increase the Higgs field value, the potential starts to be modified from the SM one around the scale $\varphi\sim M_\tx{P}/\sqrt{\xi}$ due to the change of the frames from Jordan to Einstein. Therefore we should take $\Lambda$ to be this scale (or lower) so that $\left.V\right|_{\varphi\leq\Lambda}$ is frame independent.
In the non-critical Higgs inflation, the bound~\eqref{eq:rmin} becomes weak, while in the critical one, our analysis is valid; see footnote~\ref{Higgs inflation footnote}.
}

\section{$Z_2$ Higgs-portal scalar model}\label{darkmatter}
In this paper, we employ the Higgs-portal $Z_2$ scalar dark matter.
Below the scale $\Lambda$ our Lagrangian is 
\al{
\mathcal{L}=\mathcal{L}_\text{SM}+{1\over2}(\partial_\mu S)^2-{1\over2}m_S^2 S^2-{\lambda_S\over4!}S^4-{\kappa\over2}S^2 \Phi^\dagger \Phi,
\label{lagrangian}
}
where $\Phi$ is the SM Higgs doublet and $S$ is the Higgs portal scalar field which has $Z_2$ symmetry.
Hereafter we use $\varphi:=\sqrt{2\Phi^\dagger\Phi}$.
The singlet $S$ is identified as the DM, whose mass is
\al{
m_\tx{DM}^2=m_S^2+{\kappa v^2\ov2},
}
where $v\simeq246\GeV$ is the Higgs vacuum expectation value (VEV).
The parameters $\lambda_S$ and $\kappa$ affects $V_{\varphi\leq\Lambda}$ through the renormalization group (RG) running of the Higgs quartic coupling, while $m_S$ does not.
We assume that $S$ does not acquire a Planck scale VEV and thus does not affect the inflation.

\begin{figure}[t]
  \begin{center}
   \includegraphics[height=25em]{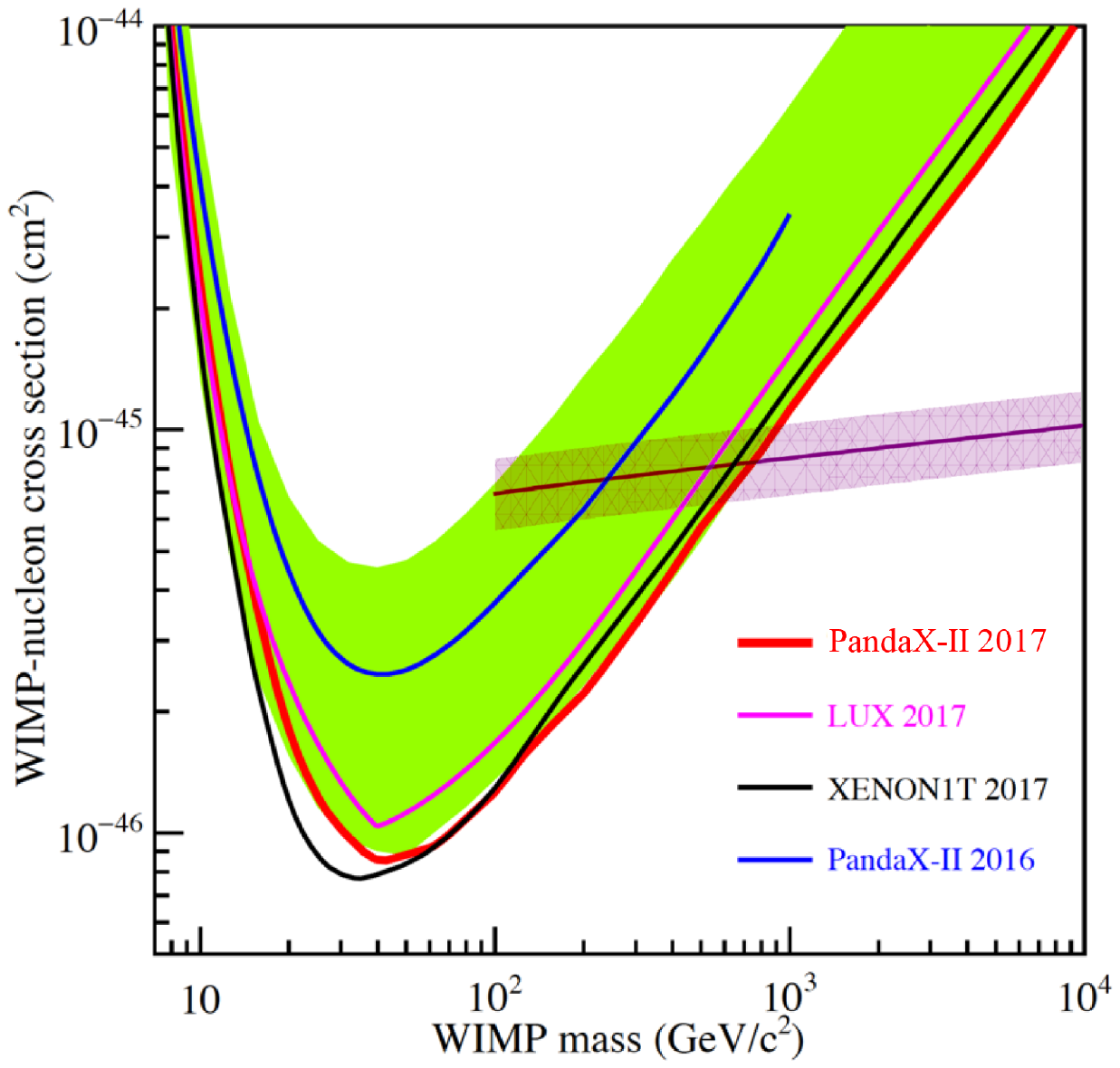}
  \end{center}
  \caption{\magenta{Fig.~5\,(a) in Ref.~\cite{Cui:2017nnn}. The upper side of each curve is excluded with $1.6\,\sigma$ C.L. We superimpose the purple line representing Eq.~\eqref{val:sigma} and the light purple region denoting the error of $\sigma_\text{SI}$ coming from the uncertainty of the overall coupling $f_N=0.30\pm0.03$; see Ref.~\cite{Cline:2013gha}.}}
  \label{fig:DM constraint}
\end{figure}

In this model, the thermal abundance of the DM fixes the relation between $\kappa$ and $m_\tx{DM}$ in the non-resonant region~\cite{Cline:2013gha}:\footnote{
In the region of our interest, $0.1\lesssim\kappa\lesssim0.5$, this roughly translates as $m_\text{DM}\simeq \kappa\times 3.2\TeV$~\cite{Hamada:2014xka}.
}
\al{
\log_{10}\kappa\simeq-3.63+1.04\log_{10}{m_\text{DM}\over\text{GeV}}.
\label{eq:convert}
}
This relation allows us to convert $m_\tx{DM}$ into $\kappa$ and vice versa.
On the other hand, the spin-independent cross section and the dark-matter mass are related by~\cite{Cline:2013gha}
\al{
\sigma_\text{SI}
    &=\frac{\kappa^2 f_N^2}{4\pi}\paren{\frac{m_n m_\tx{DM}}{m_n +m_\tx{DM}}}^2\frac{m_n^2}{m_H^4 m_\tx{DM}^2}
    \simeq 1.1\times 10^{-45} \paren{m_\tx{DM}\over\text{TeV}}^{0.08}\,\text{cm}^2.
    \label{val:sigma}
}
\magenta{
The resultant $1.6\,\sigma$ constraint on the DM mass reads, as said in Introduction,
\al{
m_\tx{DM}>0.7\pm0.2\TeV.
	\label{bound on DM mass}
}
The uncertainty $\pm0.2\,\TeV$ comes from that of the $f_N$; see the caption of Fig.~\ref{fig:DM constraint}.
Similarly, the $1.6\,\sigma$ results from the LUX~\cite{Akerib:2016vxi} and
\magenta{XENON1T~\cite{Aprile:2017iyp} experiments imply $m_\tx{DM}>0.5\pm0.1\TeV$ and $m_\tx{DM}>0.6\pm0.1\TeV$}, respectively.
The corresponding lower bound on $\kappa$ is $\kappa>0.2$ for all the three experiments. 
}

We will employ the pole mass of the top quark $m_t$ as an input parameter for the RG analysis below.
The Monte-Carlo mass of the top quark has been precisely measured to be $m_t^\tx{MC}=173.1\pm0.6\GeV$~\cite{Olive:2016xmw}.
However, the relation between $m_t^\tx{MC}$ and the pole mass $m_t$ is still unclear, and there remains uncertainty at least of 1\,GeV; see e.g.\ Ref.~\cite{Cortiana:2015rca} for a recent review.
A theorist's combination of the pole mass, derived from the cross-section measurements, reads $m_t=173.5\pm1.1\GeV$~\cite{Olive:2016xmw}.
Hereafter we take conservatively two ranges:
\al{
171\GeV<m_t&<176\GeV,	\label{2 sigma}\\
169\GeV<m_t&<178\GeV,	\label{4 sigma}
}
which roughly corresponds to $2\sigma$ and $4\sigma$ ranges of the above combination, respectively.

\clearpage


\begin{figure}[t]
  \begin{center}
   \includegraphics[height=17em]{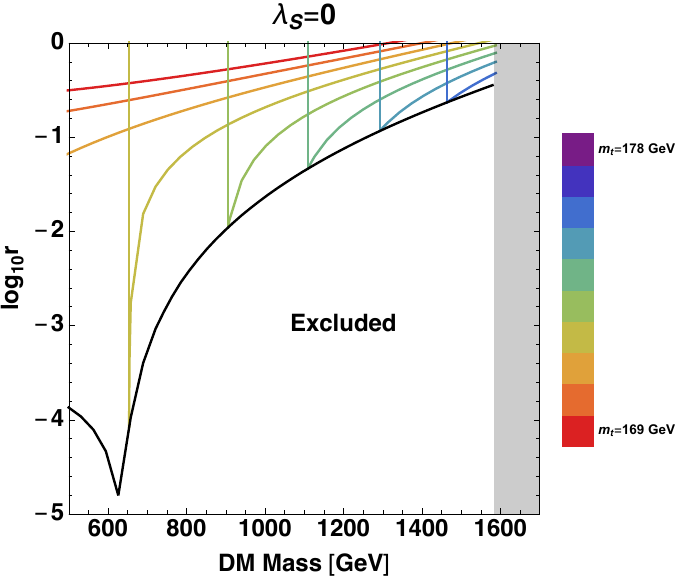}
\\
\hfill
   \includegraphics[height=17em]{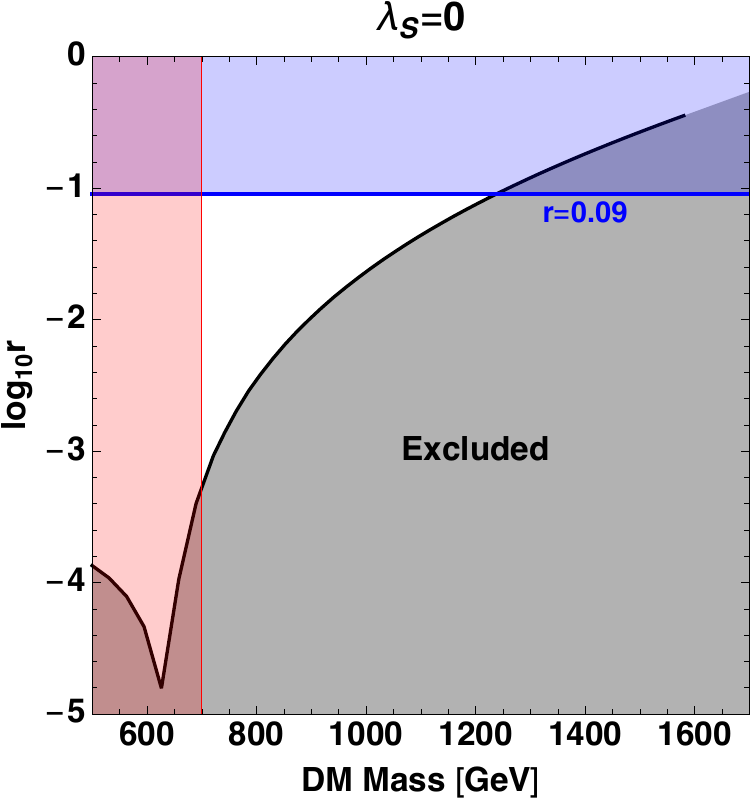}
\hfill
   \includegraphics[height=17em]{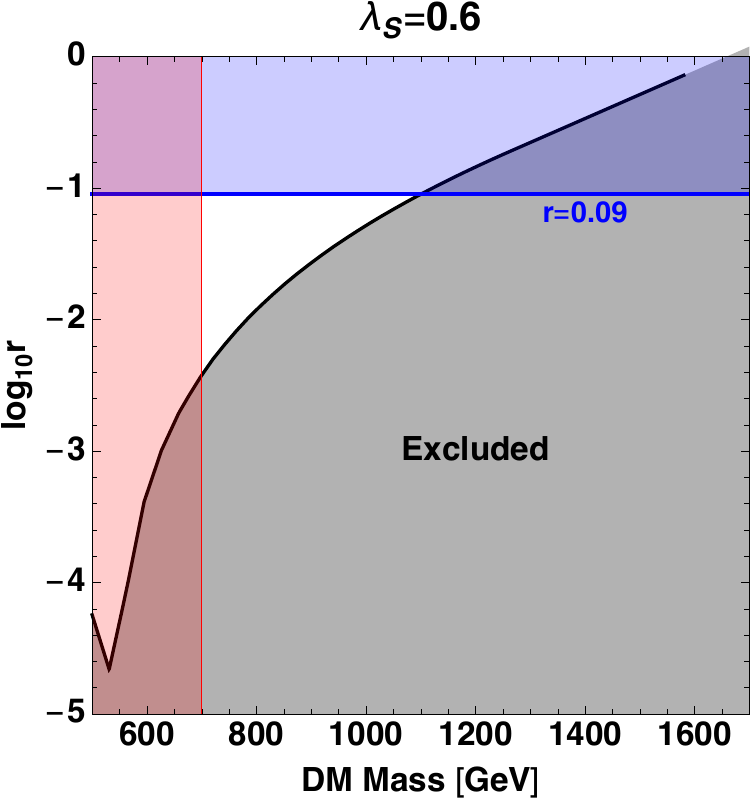}
\hfill
  \end{center}
  \caption{Upper: Allowed regions for $\lambda_S=0$. For each $m_t$, the region above the line with corresponding color is allowed. Each vertical line denotes the lower bound on $m_\tx{DM}$ from the positivity of potential: $V_{\varphi\leq\Lambda}>0$. The envelope of the rainbow-colored lines, indicated by the black line, gives the lower bound on $r$ for each $m_\tx{DM}$ when one varies $m_t$. 
  Lower: Excluded regions for $\lambda_S=0$ (lower left) and $\lambda_S=0.6$ (lower right). 
   \magenta{
  The blue and red regions are excluded by the upper bound on the tensor-to-scalar ratio $r<0.09$~\cite{Ade:2015xua} and by lower bound on the DM mass $m_\tx{DM}>0.7\TeV$ from PandaX-II~\cite{Cui:2017nnn}, respectively.}
  }
  \label{fig:N0}
\end{figure}


\section{Analysis without heavy right-handed neutrinos}\label{neutrinoless section}
As shown in the previous section, we need to compute $V_{\varphi\leq\Lambda}$ in order to put the lower bound on $r$.
We may find the excluded region in the $r$-$m_\tx{DM}$ plane by obtaining $V_{\varphi\leq\Lambda}^\tx{max}$ as a function of $\kappa$ for each fixed set of $(\lambda_S, m_t)$ and converting $\kappa$ to $m_\tx{DM}$ via Eq.~\eqref{eq:convert}.
We first present our method of analysis in Sec.~\ref{no neutrino method}.
Then we show our results in Sec.~\ref{no neutrino results}.

\subsection{Method of analysis}\label{no neutrino method}
\ef{
We solve the RGEs for various (on-shell) values of $\lambda_S$ and $\kappa$, as well as the pole mass of top quark~$m_t$. First, we input the SM parameters at $m_t$ that take into account the threshold corrections at the EW scale~\cite{Buttazzo:2013uya}. To take into account the threshold effect of the DM at the lowest order, we turn off $\lambda$ and $\kappa$ below $m_\tx{DM}$ given above.\footnote{\ef{
As all the relevant couplings are small, higher order threshold corrections are at most few percent and are beyond the scope of our consideration.
}}
More details are explained in Appendix~\ref{calculation}.
}
From the obtained running couplings, we determine $V_{\varphi\leq\Lambda}$.
Then we exclude the parameter region in which $V$ becomes negative in $\varphi\leq\Lambda$ or the perturbativity of couplings is violated. 
For the perturbativity, we demand that all the couplings are smaller than $\sqrt{4\pi}\simeq3.5$ in all the region $\varphi\leq\Lambda$. 
This condition chooses the region $\kappa\leq0.5$ ($m_\tx{DM}\lesssim1.6\TeV$) for $\lambda_S=0$.\footnote{
See Fig.~1 in Ref.~\cite{Hamada:2014xka} for the allowed region in the $\lambda_S$--$\kappa$ plane.
}
In this paper, we restrict to the case $\lambda_S=0$ except for the right of Fig.~\ref{fig:N0} in which we instead take $\lambda_S=0.6$ for comparison.\footnote{
We will see in the right of Fig.~\ref{fig:N0} that the large $\lambda_S$ tends to narrower the allowed region. Therefore, it is more conservative to set $\lambda_S=0$.
}


\subsection{Results}\label{no neutrino results}

We plot the allowed region in $r$-$m_\tx{DM}$ plane for $\lambda_S=0$ in the upper panel of Fig.~\ref{fig:N0}.
The region below each line is excluded, with its rainbow-color corresponding to each $m_t$ value.
\ef{
In the lower panels, we show a region that is excluded regardless of the top-quark mass for $\lambda_S=0$ (left) and for $\lambda_S=0.6$ (right). The shape of the excluded region depends on how the Higgs potential changes as we vary the model parameters; see Appendix~\ref{OHP} for details.
On both the panels, we also superimpose the constraints from the upper bound on $r$ (blue band) and from the lower one on $m_\tx{DM}$ (red band).
}

Fig.~\ref{fig:N0} shows that the Planck constraint $r<0.09$~\cite{Ade:2015xua} leads to bounds on $m_t$ and $m_\tx{DM}$:
\al{
171\GeV<m_t&<175 \GeV,\\
m_\tx{DM}&\lesssim1.1 \TeV.
}
This bound on $m_\tx{DM}$ is stricter than the above-mentioned perturbativity bound $m_\tx{DM}\lesssim1.6 \TeV$.

\magenta{
When we take the lower bound by the PandaX-II experiment~\cite{Cui:2017nnn}, $m_\tx{DM}\gtrsim0.7 \TeV$ ($\kappa\gtrsim0.2$), we obtain the lower bound on tensor-to-scalar ratio
\al{
r\gtrsim2\times10^{-3}.
	\label{lowest r from PandaX}
}
We can explore this possibility in near-future experiments such as the POLARBEAR-2~\cite{Inoue:2016jbg}, LiteBIRD~\cite{Matsumura:2013aja} and CORE~\cite{Delabrouille:2017rct}.
}

\clearpage
\section{Analysis with right-handed neutrinos}\label{neutrino section}

We introduce the heavy right-handed neutrinos that account for the observed neutrino masses through the seesaw mechanism, and obtain the lower bound on $r$ for each DM mass.

\subsection{Method of analysis}

\begin{table}[t]
\begin{center}
\begin{tabular}{l|cccc}
						&	$m_1$\,[eV]	&$m_2$\,[eV]	& $m_3$\,[eV]	& Pattern\\
\hline
1. Normal Hierarchy		&	0 (set)		&$8.6\times10^{-3}$&$5.1\times10^{-2}$&$m_1\ll m_2<m_3$\\
2. Inverted Hierarchy	&	$5.0\times10^{-2}$	&	$5.0\times10^{-2}$	&	0 (set)	&	$m_1\simeq m_2\gg m_3$\\
3. Degenerate (NO)		&	0.1 (set)		&$1.0\times10^{-1}$&$1.1\times10^{-1}$&$m_1\simeq m_2\simeq m_3$\\
3. Degenerate (IO)		&	$1.1\times10^{-1}$	&	$1.1\times10^{-1}$	&	0.1 (set)	&	$m_1\simeq m_2\simeq m_3$
\end{tabular}
\end{center}
\caption{Neutrino masses obtained from the absolute values of mass-squared differences in the notation of Ref.~\cite{Capozzi:2017ipn}.\label{table:nmassconst}}
\end{table}

\begin{table}[t]
\begin{center}
\begin{tabular}{l|cc}
						&	Number of effective $\nu$	& Common mass $m_\nu$ [eV]\\
\hline
1. Normal Hierarchy		&	$n_\nu=1$					& $5.1\times10^{-2}$\\
2. Inverted Hierarchy	&	$n_\nu=2$					& $5.0\times10^{-2}$\\
3. Degenerate			&	$n_\nu=3$					& $1.1\times10^{-1}$
\end{tabular}
\end{center}
\caption{Common neutrino mass that we use as an input.\label{table:nmass}}
\end{table}

The observational constraints on mass of left-handed neutrinos are the upper bound on the sum of masses and their squared differences; see e.g.\ Ref.~\cite{Capozzi:2017ipn}.
Under this condition, we consider the following three typical patterns of mass relations:
\begin{enumerate}
\item Normal Hierarchy (NH, $m_1$ the lightest),
\item Inverted Hierarchy (IH, $m_3$ the lightest),
\item Degenerate (all masses comparable).
\end{enumerate}
The mass pattern is most hierarchical when the lightest one is 0.
In Degenerate case, the upper bound on the sum of neutrino masses reads $m_i\lesssim0.1\eV$.\footnote{
The left-handed neutrino mass $0.1\eV$ for the three degenerate neutrinos corresponds to $\sum_i m_i=0.3\eV$.
The $2\sigma$ upper bound from the TT-only analysis is $\sum_i m_i<0.715\eV$, while that from the TT+lensing+ext gives $\sum_i m_i<0.234\eV$~\cite{Ade:2015xua}.
See also Refs.~\cite{Giusarma:2016phn,Vagnozzi:2017ovm} for more recent analyses that give a tighter bound $\sum_im_i<0.12\,\text{eV}$.
}
In Table~\ref{table:nmassconst}, we show the mass pattern by setting the lightest one to be zero (0.1\,eV) for the cases of normal/Inverted Hierarchy (Degenerate), using the mass-squared differences in Ref.~\cite{Capozzi:2017ipn}.
For the three cases, we approximate the heaviest $n_\nu$ neutrinos as having a common mass $m_\nu$ and the remaining $3-n_\nu$ ones as being massless as shown in Table~\ref{table:nmass}.\footnote{
If we want to consider a different $m_\nu$, we may simply rescale the right-handed neutrino mass $\MR$ in our results, since $m_\nu\propto \MR^{-1}$ by the seesaw mechanism.
}

Under the existence of heavy right-handed neutrino, the remaining input parameters to determine $V_{\varphi\leq\Lambda}$ are $\lambda_S$, $\kappa$, $m_t$, and the right-handed neutrino mass~$M_{\tx{R},i}$.
For simplicity, we assume that $M_{\tx{R},i}$ ($i=1,2,3$) are identical: $M_{\tx{R},i}=\MR$.
The Yukawa coupling of neutrino is given by the seesaw mechanism: $y_\nu=\sqrt{2m_\nu \MR}/v$, with $v\simeq246\GeV$.
\ef{
The threshold effect is taken into account at the lowest order by turning on $y_\nu$ above $\MR$ in the RGEs.\footnote{\ef{
Higher order threshold corrections are beyond the scope of our consideration; see footnote~\ref{roughness}.
}}
See Appendix~\ref{calculation} for more details.
}


\subsection{Results for Normal Hierarchy}\label{sec:result normal}

\begin{figure}[p]
  \begin{center}
  \vspace{-1.0em}
   \includegraphics[height=17em]{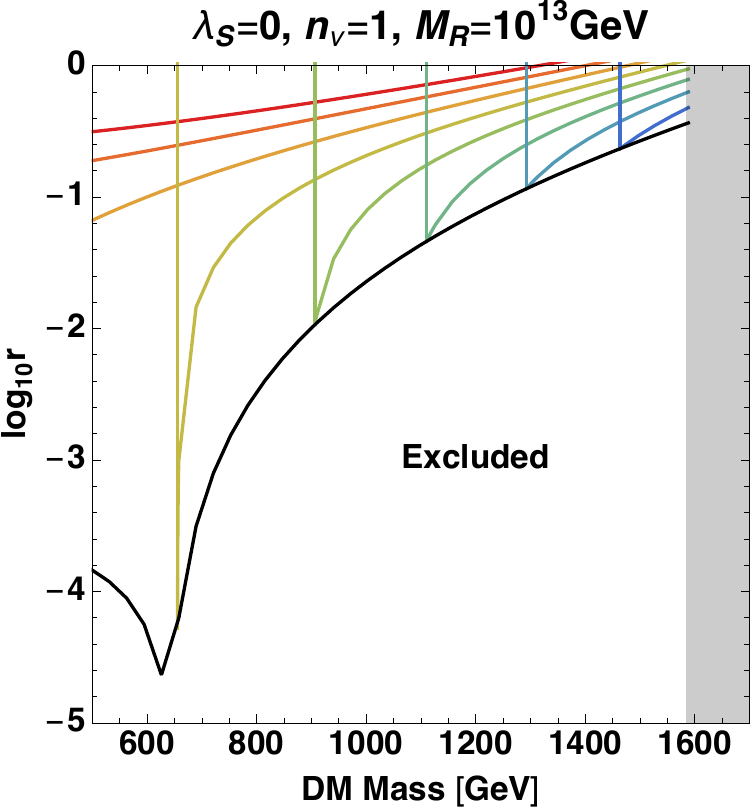}
 \hfill
   \includegraphics[height=17em]{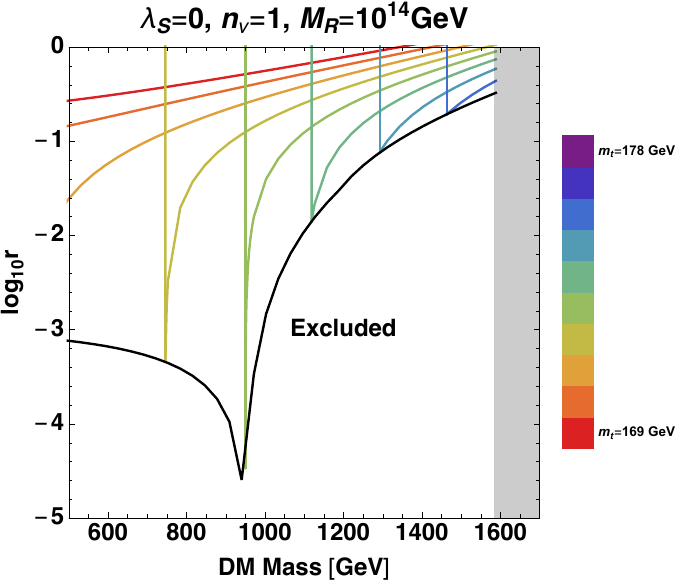}
  \end{center}
  \begin{center}
   \includegraphics[height=17em]{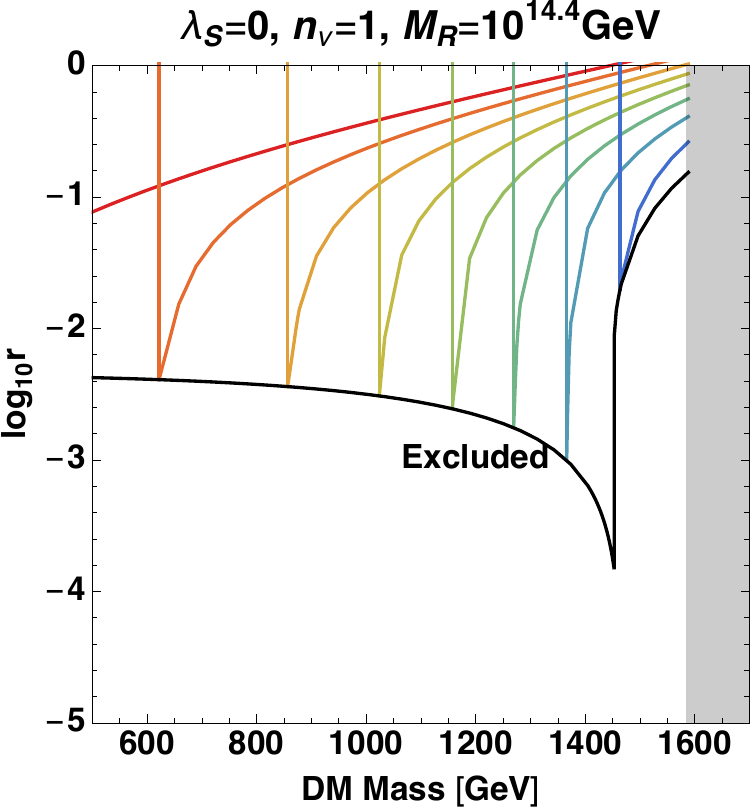}
 \hfill
   \includegraphics[height=17em]{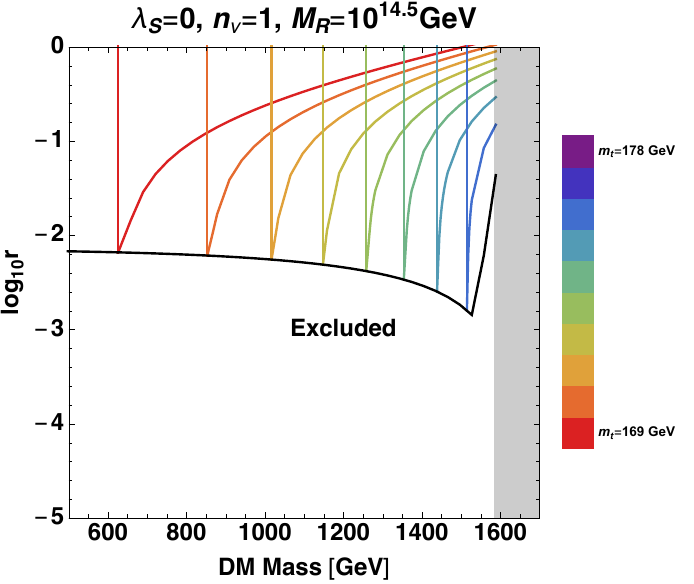}
  \end{center}
  \begin{center}
   \includegraphics[height=17em]{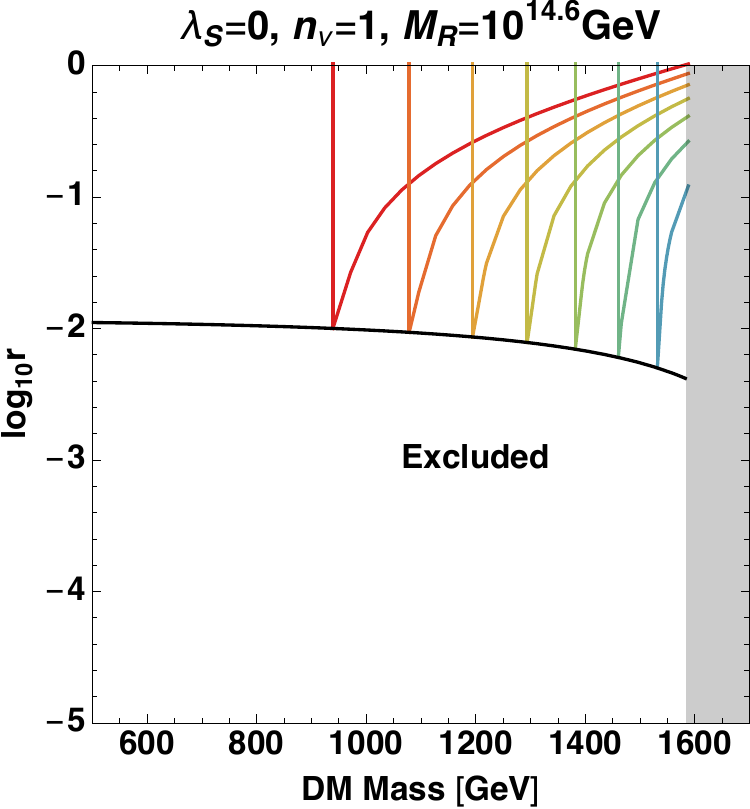}
 \hfill
   \includegraphics[height=17em]{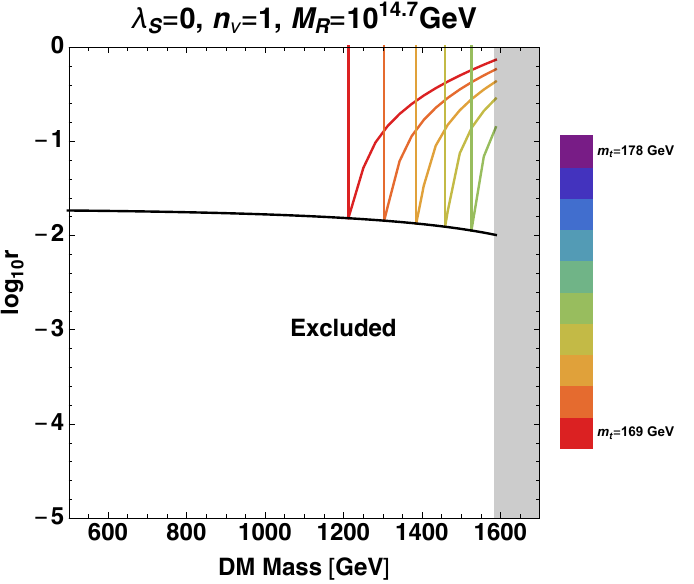}
  \end{center}
  \caption{Allowed region for Normal Hierarchy with $\lambda_S=0$. The right-handed neutrino mass $\MR$ is increased in the order of the upper-left, upper-right, middle-left, \dots, lower-right panels. The bold line in each panel is the envelope of the $m_t$-fixed rainbow-colored lines, and gives the lower bound on $r$ for the fixed $\MR$. See the caption of Fig.~\ref{fig:N0} for the shaded region.}
  \label{fig:N1}
\end{figure}

\begin{table}[t]
\begin{center}
\begin{tabular}{l|cc}
								& $m_t$					& $\MR$\\
\hline
$m_\tx{DM}=1\TeV$, $r=0.01$		& $m_t<174\GeV$			& $10^{14}\GeV\lesssim \MR\lesssim10^{14.6}\GeV$\\
$m_\tx{DM}=1\TeV$, $r=0.001$	& $173\GeV<m_t<174\GeV$	& $10^{14.1}\GeV\lesssim \MR\lesssim10^{14.3}\GeV$\\
$m_\tx{DM}=1.5\TeV$, $r=0.01$	& $170\GeV<m_t<178\GeV$	& $10^{14.6}\GeV\lesssim \MR\lesssim10^{14.8}\GeV$\\
$m_\tx{DM}=1.5\TeV$, $r=0.001$	& $m_t\simeq177.8\GeV$	& $\MR\simeq10^{14.6}\GeV$
\end{tabular}
\end{center}
\caption{Constraints that will be obtained from future observations of $m_\tx{DM}$ and $r$ for Normal Hierarchy.\label{table:result for NH}}
\end{table}

We show the results for Normal Hierarchy, $n_\nu=1$, in Fig.~\ref{fig:N1}.
The right-handed neutrino mass $\MR$ is fixed in each panel: $10^{13}$, $10^{14}$, $10^{14.4}$\,($\simeq2.5\times10^{14}$), $10^{14.5}$\,($\simeq3.2\times10^{14}$), $10^{14.6}$\,($\simeq4.0\times10^{14}$), and $10^{14.7}$\,($\simeq5.0\times10^{14}$) in units of GeV.
The color of envelope in each panel, denoted by the thick line, corresponds to the color in the plots in Sec.~\ref{sec:all},
in which the discussion for more general values of $\MR$ will be given.
Note that the thick line is obtained by tuning one parameter $m_t$ for fixed $\MR$, and its minimum corresponds to the two parameter tuning of $m_t$ and $m_\tx{DM}$.

With the right-handed neutrino, we have one more theoretical parameter $\MR$ in addition to $m_t$ to determine from the observational constraints of $r$ and $m_\tx{DM}$.
From Fig.~\ref{fig:N1}, we see that the larger the $\MR$ is, the smaller the allowed region becomes, for a given lower bound on $m_t$.
This is because the right-handed neutrinos and the top quark have a similar effect on the Higgs potential, namely, they drives the Higgs quartic coupling smaller through the RG running towards high scales, and therefore they tend to make the Higgs potential negative if they both are heavy.\footnote{
In this case, the vacuum stability is violated before the perturbativity.
}
If we e.g.\ set $m_t>169\GeV$, we have the constraint $\MR\lesssim10^{14.8}\GeV$ ($\simeq6.3\times10^{14}\GeV$); see also Fig.~\ref{fig:houraku4sigma} in Sec.~\ref{sec:all}.

As we increase $\MR$ and switch panels in Fig.~\ref{fig:N1}, we see that the value of $m_\tx{DM}$ at the minimum point of the envelope becomes larger: 600\GeV, 870\GeV, etc. In particular, it goes beyond the perturbativity bound, indicated by the gray band, when  $\MR=10^{14.7}\GeV$.
Therefore, if the right-handed neutrino mass is larger than that, we have a stringent lower bound: $r\gtrsim 10^{-2}$.
On the other hand, the plot with $\MR\lesssim10^{13}\GeV$ is almost the same as the case without right-handed neutrinos shown in the left of Fig.~\ref{fig:N0}.

Let us see implications of future discoveries of the DM and $r$ \ef{on the right-handed neutrino mass in the current context}:\footnote{
Here, we fit $\MR$ and $m_t$ from the future observation of $m_\tx{DM}$ and $r$.
Instead, one might narrow down the error on the pole mass $m_t$ e.g.\ at the High Luminosity LHC (HL-LHC)~\cite{CMS-DP-2016-064}.
Then one may use $m_t$ and $m_\tx{DM}$ as input parameters to predict $\MR$ and $r$.
}
\begin{itemize}
\item Suppose that $m_\tx{DM}=1\TeV$ ($\kappa\simeq0.31$) and $r=0.01$ are found.
Then the right-handed neutrino mass is predicted to be in the narrow range $10^{14}\GeV\lesssim \MR\lesssim10^{14.6}\GeV$ and the top-quark mass is constrained from above: $m_t<174\GeV$.
\item If we discover $m_\tx{DM}=1.5\TeV$ ($\kappa\simeq0.47$) and $r=0.01$, we obtain the theoretical lower bound $\MR\gtrsim10^{14.6}\GeV$, while the top quark mass is less constrained: $171\GeV<m_t<178\GeV$.
However, $\MR$ and $m_t$ are highly correlated in this case.
Therefore if one of them is fixed, the other is precisely predicted.
\item See Table~\ref{table:result for NH} for other pairs of $m_\tx{{DM}}$ and $r$. Generically the heavy DM mass tends to predict the heavy top-quark mass and $\MR$. The smaller the $r$ is, the tighter the range of $m_t$. Especially, if we discover $m_\tx{DM}=1.5\TeV$ and $r=0.001$, $m_t$ and $\MR$ are accurately predicted.
\end{itemize}

We can predict $r_\tx{bound}$ or $m_\tx{DM}$ to some extent by considering typical input parameters.
When we choose $m_t=173\GeV$ and $\MR=10^{14}\GeV$, we obtain the bound $m_\tx{DM}\sim860\GeV$--$970\GeV$ for $r<0.09$.

\subsection{Results for Inverted Hierarchy}\label{sec:result inverted}

\begin{figure}[t]
  \begin{center}
   \includegraphics[height=17em]{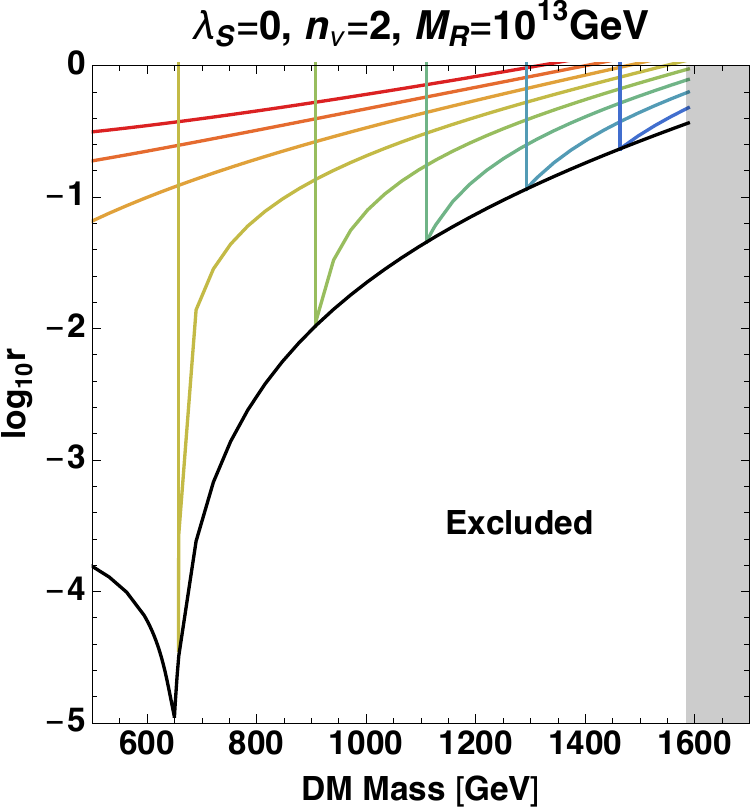}
 \hfill
   \includegraphics[height=17em]{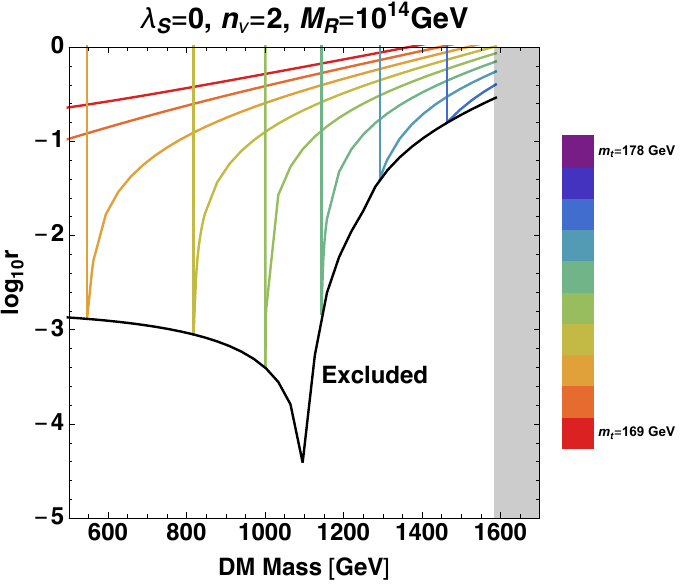}
  \end{center}
  \begin{center}
   \includegraphics[height=17em]{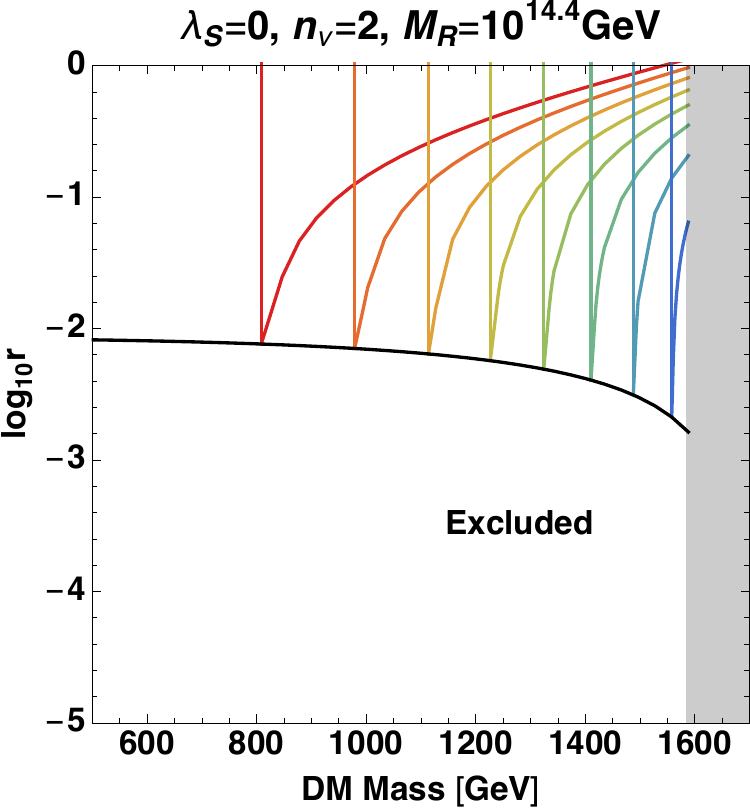}
 \hfill
   \includegraphics[height=17em]{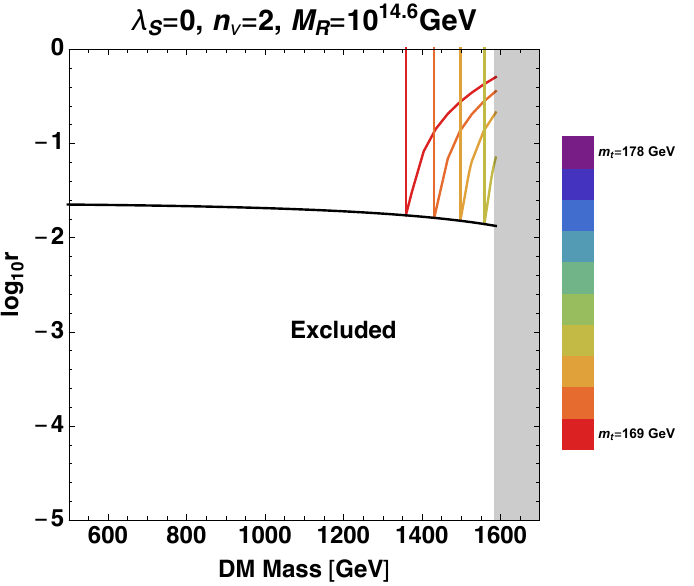}
  \end{center}
  \caption{Allowed region for Inverted Hierarchy with $\lambda_S=0$.
  See caption of Fig.~\ref{fig:N1} for the explanation.
  }
  \label{fig:N2}
\end{figure}

We show the results for the case of Inverted Hierarchy ($n_\nu=2$) in Fig.~\ref{fig:N2}.
In this case, the right-handed neutrinos lighter than $\sim10^{13}\GeV$ do not affect the analysis, similarly as in the case of Normal Hierarchy.
However, the upper bound on $\MR$ is slightly different: $\MR\lesssim10^{14.7}\GeV$, see also Fig.~\ref{fig:houraku2sigma}.
Let us summarize implications of future discoveries of the DM and $r$ again \ef{on the right-handed neutrino mass in the current context}:
\begin{itemize}
\item If we discover $m_\tx{DM}=1\TeV$ and $r=0.01$, we obtain $10^{13.9}\GeV\lesssim \MR\lesssim10^{14.4}\GeV$ and $m_t<174\GeV$.
\item If we discover $m_\tx{DM}=1.5\TeV$ and $r=0.01$, $\MR$ must be larger than $\sim10^{14.6}\GeV$ and $170\GeV<m_t<178\GeV$.
Although we cannot obtain the global narrow bounds on $\MR$ and $m_t$, they are highly correlated as in the case of Normal Hierarchy.
\end{itemize}
See Table~\ref{table:result for IH} for other pairs of $m_\tx{{DM}}$ and $r$.

\clearpage

\begin{table}[h]
\begin{center}
\begin{tabular}{l|cc}
								& $m_t$					& $\MR$\\
\hline
$m_\tx{DM}=1\TeV$, $r=0.01$		& $m_t<174\GeV$			& $10^{13.9}\GeV\lesssim \MR<10^{14.5}\GeV$\\
$m_\tx{DM}=1\TeV$, $r=0.001$	& $173\GeV<m_t<174\GeV$	& $10^{13.9}\GeV<\MR<10^{14.2}\GeV$\\
$m_\tx{DM}=1.5\TeV$, $r=0.01$	& $m_t<178\GeV$			& $10^{14.4}\GeV<\MR\lesssim10^{14.7}\GeV$\\
$m_\tx{DM}=1.5\TeV$, $r=0.001$	& $177\GeV<m_t<178\GeV$	& $10^{14.4}\GeV<\MR<10^{14.5}\GeV$
\end{tabular}
\end{center}
\caption{Constraints obtained for Inverted Hierarchy.\label{table:result for IH}}
\end{table}

\clearpage

\subsection{Results for Degenerate case}\label{sec:result degenerate}
\begin{figure}[h]
  \begin{center}
   \includegraphics[height=17em]{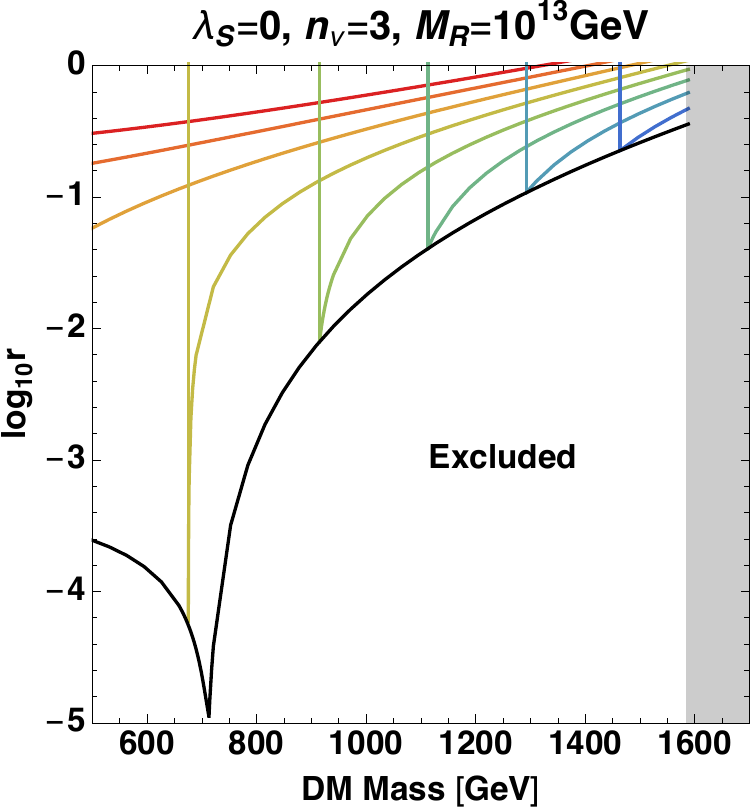}
\hfill
   \includegraphics[height=17em]{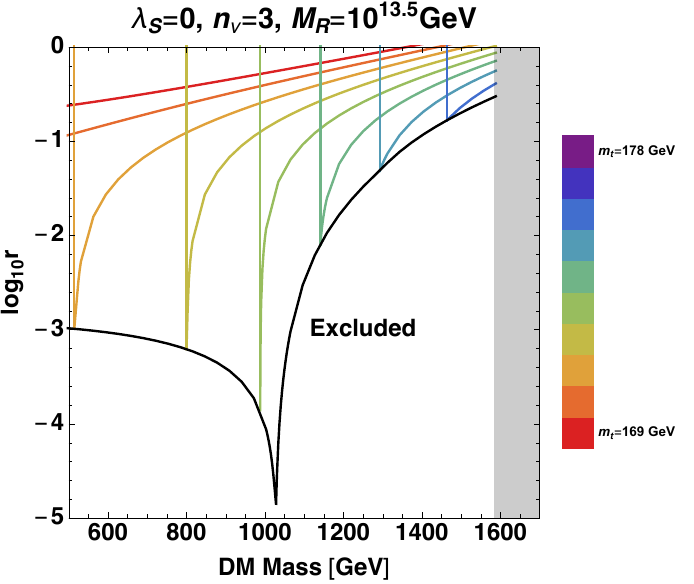}
  \end{center}
  \begin{center}
   \includegraphics[height=17em]{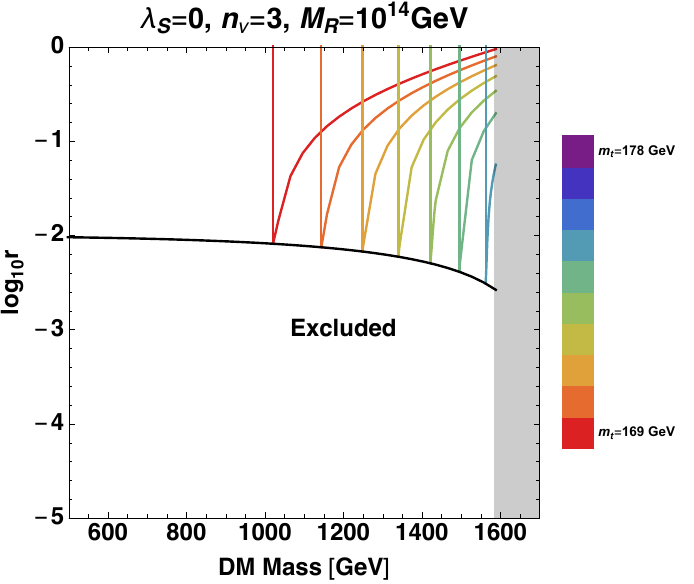}
  \end{center}
  \caption{Allowed region for Degenerate case with $\lambda_S=0$.
  See caption of Fig.~\ref{fig:N1} for the explanation.
  }
  \label{fig:N3}
\end{figure}

We show the results for Degenerated case ($n_\nu=3$) in Fig.~\ref{fig:N3}.
The right-handed neutrinos lighter than $\sim10^{13}\GeV$ do not affect the analysis, similarly as other cases.
The upper bound on $\MR$ is smaller than in other cases: $\MR\lesssim10^{14.2}\GeV\simeq1.6\times10^{14}\GeV$.

We summarize implications of future discoveries $m_\tx{DM}$ and $r$ in Table~\ref{table:result for DC}.
The right-handed neutrino mass tend to be lighter than hierarchical cases due to the heavy $m_\nu$.
However, the prediction of $m_t$ is similar to the other cases.

\clearpage

\begin{table}[t]
\begin{center}
\begin{tabular}{l|cc}
								& $m_t$					& $\MR$\\
\hline
$m_\tx{DM}=1\TeV$, $r=0.01$		& $m_t<174\GeV$					& $10^{13.5}\GeV\lesssim \MR\lesssim10^{14}\GeV$\\
$m_\tx{DM}=1\TeV$, $r=0.001$	& $173\GeV\lesssim m_t<174\GeV$	& $10^{13.5}\GeV<\MR<10^{13.8}\GeV$\\
$m_\tx{DM}=1.5\TeV$, $r=0.01$	& $m_t<178\GeV$					& $10^{14}\GeV\lesssim\MR\lesssim10^{14.2}\GeV$\\
$m_\tx{DM}=1.5\TeV$, $r=0.001$	& $176\GeV<m_t<178\GeV$			& $\MR\simeq10^{14}\GeV$
\end{tabular}
\end{center}
\caption{Constraints obtained for degenerate case.\label{table:result for DC}}
\end{table}

\section{Allowed region for all parameter space}\label{sec:all}
In the previous section, we have plotted the bounds for various $m_t$ in each panel of fixed $\MR$. In each panel, we have also shown the envelope of different $m_t$ lines.
This envelope is our theoretical lower bound on $r$ for a given $\MR$.

In Sec.~\ref{envelope of envelope for fixed MR}, we show these envelopes altogether in the same plot, and give the absolute lower bound on $r$ for varying $m_t$ and $\MR$.
In Sec.~\ref{envelope of envelope for fixed mt}, we see the same absolute lower bound in a different way, by changing the order of fixing $m_t$ and $\MR$.

\subsection{Lower bound on $r$ for each $\MR$}\label{envelope of envelope for fixed MR}

\begin{figure}[t]
  \begin{center}
   \includegraphics[height=17em]{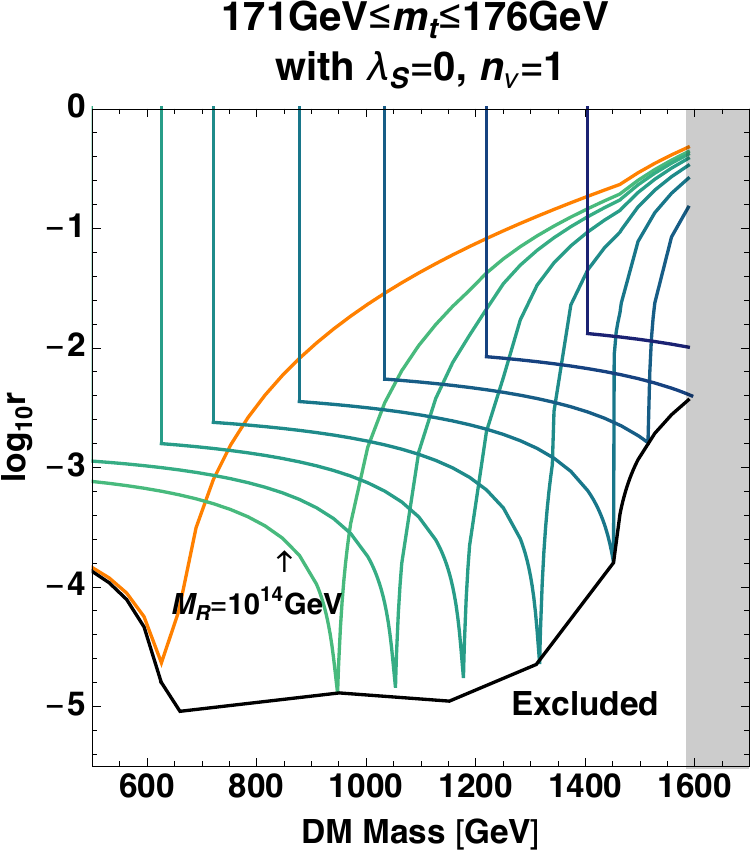}
 \hfill
   \includegraphics[height=17em]{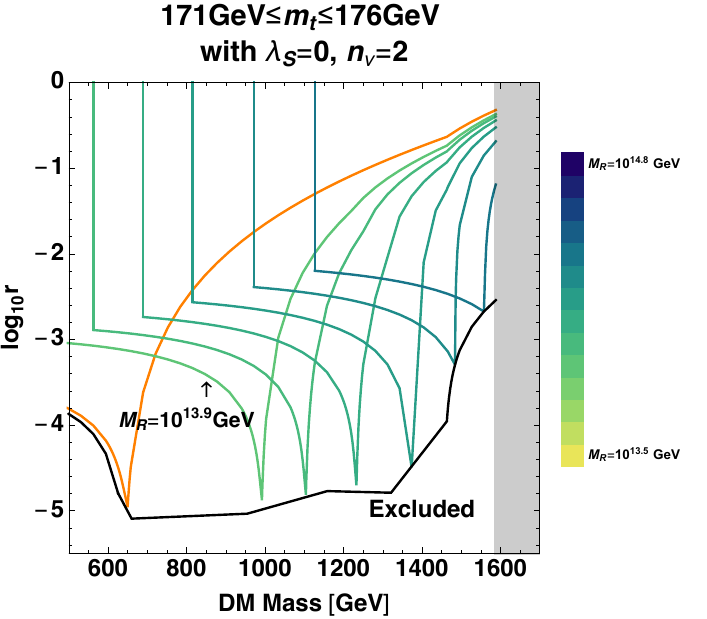}
  \end{center}
  \begin{center}
   \includegraphics[height=17em]{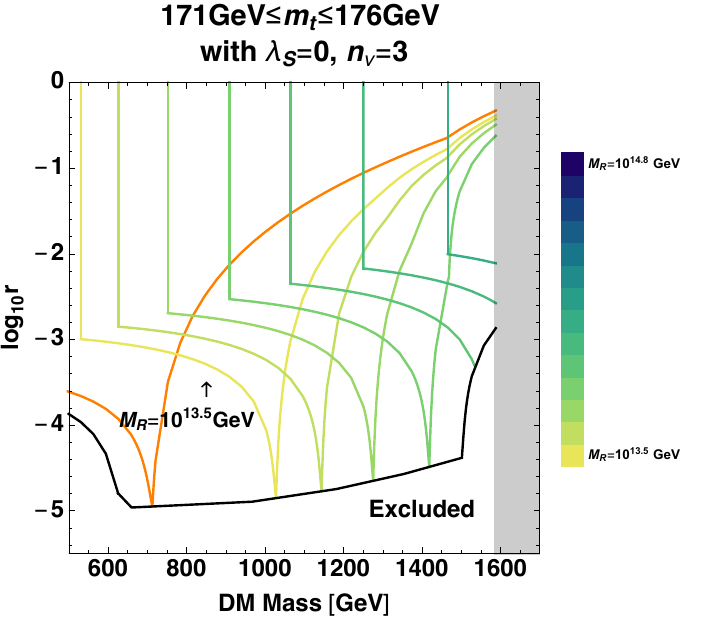}
  \end{center}
  \caption{The lower bound on $r$ for each fixed $\MR$ (colored) and the envelope (black) with $\lambda_S=0$ and $171\GeV< m_t<176\GeV$. The orange line is for $\MR=10^{13}$ GeV. The vertical colored line comes from the lower end, $m_t>171\GeV$.
  See the caption of Fig.~\ref{fig:N0} for the shaded region.}
  \label{fig:houraku2sigma}
\end{figure}

\begin{figure}[t]
  \begin{center}
   \includegraphics[height=17em]{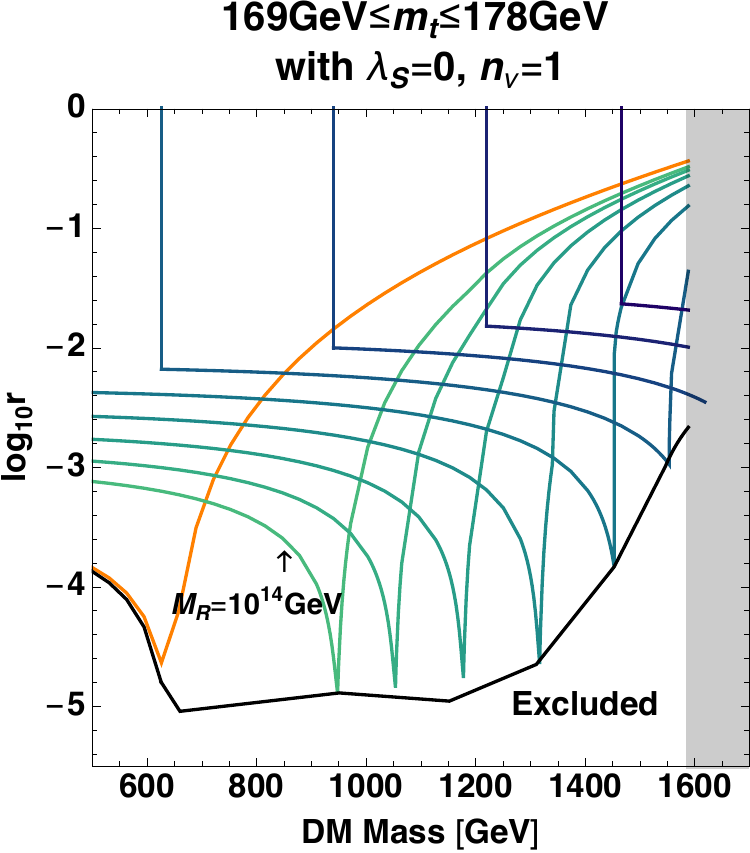}
\hfill
   \includegraphics[height=17em]{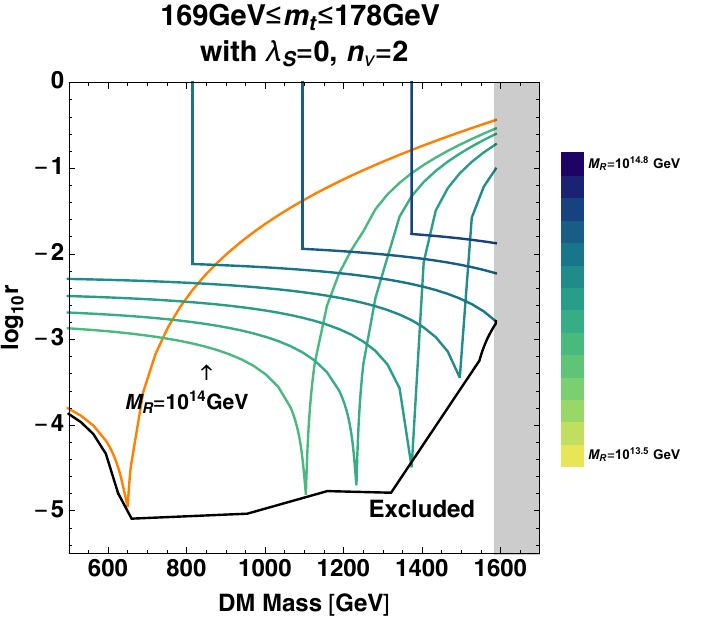}
  \end{center}
  \begin{center}
   \includegraphics[height=17em]{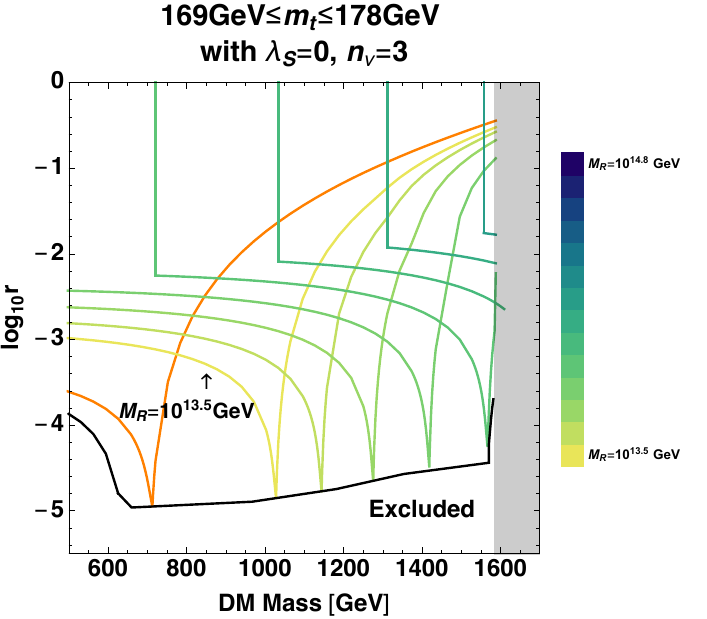}
  \end{center}
  \caption{The same plot as in Fig.~\ref{fig:houraku2sigma} for $169\GeV<m_t<178\GeV$.}
  \label{fig:houraku4sigma}
\end{figure}

In Figs.~\ref{fig:houraku2sigma} and \ref{fig:houraku4sigma}, we plot our theoretical lower bounds on $r$ for various $\MR$ when we allow the top-quark pole mass within roughly $2\sigma$ and $4\sigma$ ranges shown in Eqs.~\eqref{2 sigma} and \eqref{4 sigma}, respectively.
We also give the envelope of these lines, which gives the allowed region for varying $m_t$ and $\MR$.
In the plot, each colored line represents the lower bound on $r$, and corresponds to the envelope denoted by the thick colored line in Secs.~\ref{sec:result normal}--\ref{sec:result degenerate}.
We also show the absolute lower bound by the black line.\footnote{
We have plotted the envelope, denoted by the black line, as follows:
1) Each $\MR$-fixed line has a minimum. Make an interpolating function which linearly join all these minimum points. 
2) Each $m_t$-fixed line has a minimum. Make another interpolating function which linearly join all these minimum points.
3) Make a function that chooses the smaller value of these two for each $m_\tx{DM}$.
4) In large $m_\tx{DM}$ region, we replace the interpolated bound with the lower bound determined by the maximal $m_t$; see the caption of Fig.~\ref{fig:mtfix-main} to see how $m_t$ gives the bound.
Note that these interpolating functions are evaluated only for $600\GeV<m_\tx{DM}<1600\GeV$, and hence they are untrustworthy in the extrapolated regions $m_\tx{DM}<600\GeV$ and $m_\tx{DM}>1600\GeV$. This does not do any harm because these regions are already excluded by the direct DM search and by the perturbativity, respectively.
}

We explain the envelope denoted by the black line in Fig.~\ref{fig:houraku2sigma}:
\begin{itemize}
\item We see that the allowed region is enlarged to
\al{
r\gtrsim10^{-5}
}
from the $n_\nu=0$ case in Eq.~\eqref{lowest r from PandaX}, which is read from the black line in Fig.~\ref{fig:N0} (being close the orange $\MR=10^{13}\GeV$ line in Fig.~\ref{fig:houraku2sigma}).
This is because the loop corrections of heavy right-handed neutrinos reduce $V_{\varphi\leq\Lambda}^\tx{max}$.
\item The lower bound on $r$ increases rapidly in the region $m_\tx{DM}\gtrsim1.3\TeV$ due to the upper end of the parameter $m_t<176\GeV$.
\item In the region near the envelope denoted by the black line, the two input parameters $m_t$ and $\MR$ are simultaneously tuned to minimize the potential height $V_{\varphi\leq\Lambda}^\tx{max}$.
\item If one allows to adjust the three parameters, $m_t$, $\MR$, and $m_\tx{DM}$ simultaneously, then the lowest point of the black line, $r\sim10^{-5}$, is realized.
This might be the case if some logic that demands the fine tuning, such as the multiple-point principle~\cite{Froggatt:1995rt,Froggatt:2001pa,Nielsen:2012pu}, is indeed applicable.
\end{itemize}

In Fig.~\ref{fig:houraku4sigma}, we plot for a wider range of the top-quark mass~\eqref{4 sigma}.
The lower bound on $r$, denoted by the black line, increases in the region $m_\tx{DM}\gtrsim 1.3\TeV$ in the cases of $n_\nu=1$ and 2 because of the difference of potential shapes explained in Appendix~\ref{OHP}, while its rise in the region $m_\tx{DM}\gtrsim1.5\TeV$ is due to the upper end of the parameter $m_t<178\GeV$.


\subsection{Lower bound on $r$ for each $m_t$}
\label{envelope of envelope for fixed mt}

\begin{figure}[t]
  \begin{center}
   \includegraphics[height=17em]{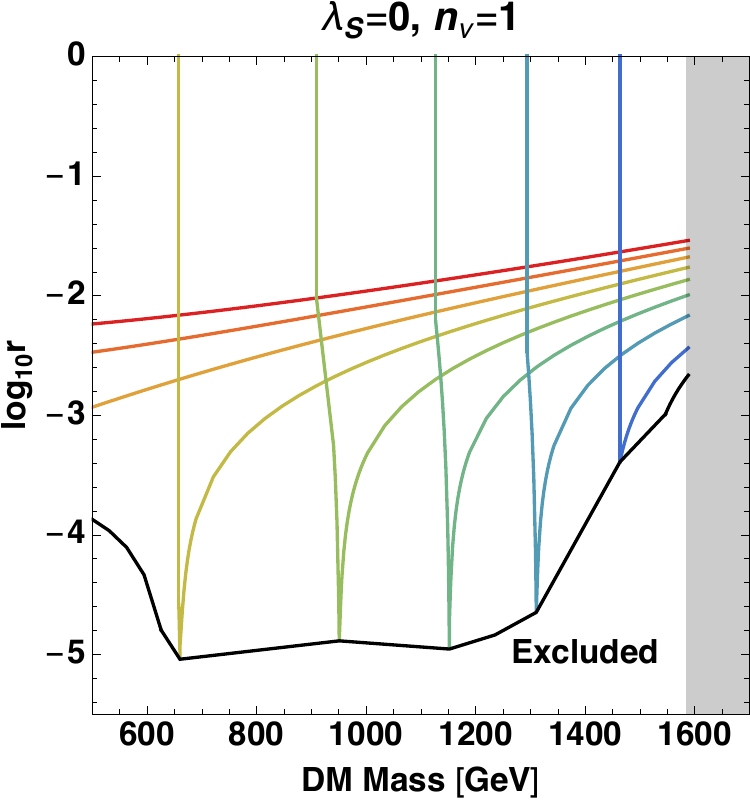}
\hfill
   \includegraphics[height=17em]{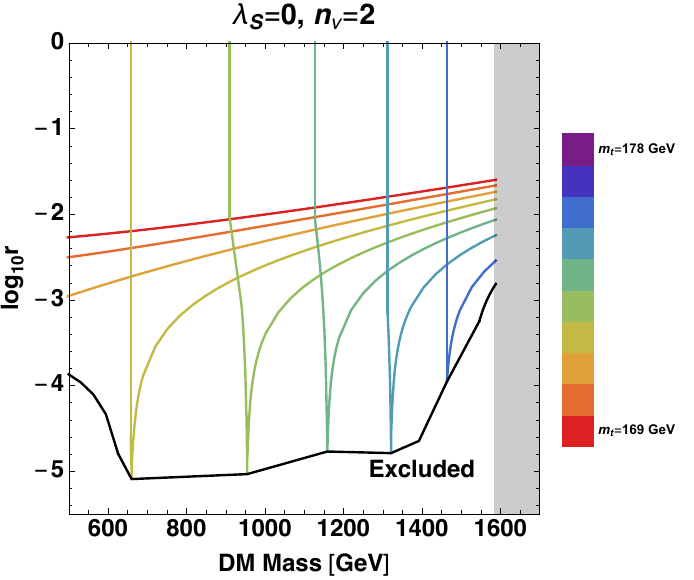}
  \end{center}
  \begin{center}
   \includegraphics[height=17em]{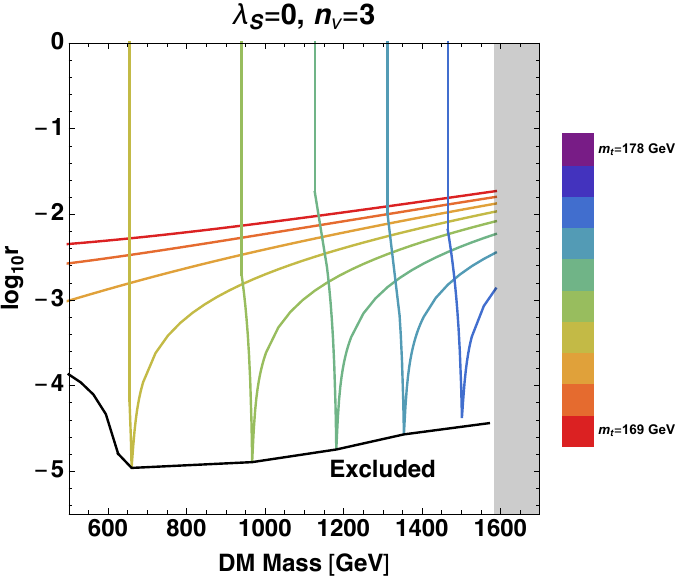}
  \end{center}
\caption{The lower bound on $r$ for each fixed $m_t$ (colored) with $\lambda_S=0$.
The black line is their envelope, which is identical to the ones in Figs.~\ref{fig:houraku2sigma} and \ref{fig:houraku4sigma} except for their right-most boundary where they follow the $m_t=176\GeV$ (blue) and 178\,GeV (purple) lines in this figure, respectively.
See the left of Fig.~\ref{fig:N0} for the corresponding plot without right-handed neutrinos and for the explanation of the shaded region.
}
\label{fig:mtfix-main}
\end{figure}

In Fig.~\ref{fig:mtfix-main}, we show the lower bound on $r$ for each fixed $m_t$ with $\MR$ being varied.
The envelope, denoted by the black line, is the same as the global lower bound on $r$ in Sec.~\ref{envelope of envelope for fixed MR} except for the range explained in the caption.

The lowest possible value of $m_\tx{DM}$ for each $m_t$ does not depend on the number of neutrinos~$n_\nu$. Therefore, for any given lower bound on $m_t$, we obtain the corresponding lower bound of $m_\tx{DM}$ without any assumptions on the other parameters in the neutrino sector.
We see that the lowest point for a given $m_t$ moves right as we increase $m_t$.
This leads to a strong correlation between $m_t$ and $m_\tx{DM}$ regardless of $n_\nu$ if $r<10^{-3}$.

\clearpage

\section{Summary and discussion}\label{summary section}

We have calculated the lower bound on the tensor-to-scalar ratio, $r$, for each given mass of the Higgs-portal $Z_2$ scalar dark matter, $m_\tx{DM}$, under the simple assumption that the extrapolation of the Higgs-field direction plays the role of inflaton at $\varphi>\Lambda$.
The advantage of our approach is that we may obtain the lower bound on $r$ without knowing any detail of the high-scale physics.

In the case without the heavy right-handed neutrinos, we have obtained the theoretical bounds on (i)~the DM mass $m_\tx{DM}\lesssim1.1\TeV$, (ii)~the tensor-to-scalar ratio $r\gtrsim\magenta{2}\times10^{-3}$, and (iii)~the pole mass of the top quark $171\GeV<m_t<175\GeV$ from the current observational constraints $r<0.09$ and $m_\tx{DM}\gtrsim\magenta{0.7\TeV}$.
We see that (i) and (ii) are rather stringent and are well within the near-future detection.

With the heavy right-handed neutrinos, we obtain the wider allowed region in the $r$-$m_\tx{DM}$ plane, $r\gtrsim10^{-5}$ and $m_\tx{DM}\lesssim1.6\TeV$, if we allow a three-parameter tuning.
Altough the region $r\lesssim10^{-3}$ is hard for the planned near-future observations, we may still explore it in combination with the HL-LHC and future neutrino experiments because of the strong correlations between $m_t$, $m_\tx{DM}$ and the right-handed neutrino mass $\MR$.

The lower bound on $r$ may slightly be affected when we relax the positivity condition on the Higgs potential by e.g.\ taking into account the thermal correction or by replacing it with the vacuum meta-stability.
Because our bound is coming from the maximum value of effective Higgs potential, rather than the minimum, the lower bound on $r$ would be reduced only by a factor of few even if we allow the negative value of the potential minimum of the order of the height of the potential maximum.
Of course we should make sure that finally the EW vacuum is chosen in the late time in such a case.

In our RG analysis, we have assumed that all the fields are massless.\footnote{
The right-handed neutrinos are regarded massless at $\mu\sim\varphi_\tx{cl}\gtrsim\MR$.
}
In general, the non-minimal couplings $\xi\varphi^2R$ and $\xi_SS^2R$ cause the mass terms of the order of $\xi H^2\varphi^2$ and $\xi_SH^2S^2$, respectively.
Under a classical field value $\varphi_\tx{cl}$, we get
\al{
\xi H^2\varphi^2&\sim \xi\lambda{\varphi_\tx{cl}^4\ov\MP^2}\varphi^2, &
\xi_S H^2S^2&\sim \xi_S\lambda{\varphi_\tx{cl}^4\ov\MP^2}S^2.
}
Such effective masses $\sqrt{\xi\lambda}{\varphi_\tx{cl}^2/\MP}$ and $\sqrt{\xi_S\lambda}{\varphi_\tx{cl}^2/\MP}$ are smaller enough than $\mu\sim\varphi_\tx{cl}$ since we consider the  non-minimal couplings $\xi,\xi_S\lesssim10^2$ and the small quartic coupling $\lambda\ll 10^{-1}$ in the region $\varphi_\tx{cl}\lesssim10^{17}\GeV$.

It would be interesting to investigate the cosmology of our scenario after the inflation.
For example, the Higgs field may be trapped at a false vacuum, and a mini inflation may results from it, depending on the initial condition at the end of the main inflation.\footnote{
In this case, the dynamics of the Higgs field becomes chaotic in the sense that it is sensitive to the initial condition at the end of the main inflation.
}
Afterwards, the true-vacuum bubbles should be created by the tunneling process, and the first order phase transition be completed by the bubble collision, which generates primordial gravitational waves and black holes.
In addition, if the Higgs potential has an inflection point, the scalar perturbations could be enhanced, depending on how the slow-roll condition is well satisfied there, and could lead to another mechanism of the formation of primordial black holes.
We hope to return to these issues in the future.

\blue{
\paragraph{Note added:}
After submission of the manuscript, there appeared newer
bound~\cite{Aprile:2018dbl}, which roughly raise the lower bound on the dark matter
mass up to $m_\text{DM}>0.9\pm0.2$\,TeV from Eq.~\eqref{bound on DM mass}.
}

\section*{Acknowledgement}
We thank Tomohiro Abe, Xiangdong Ji, Kiyoharu Kawana, Jinsu Kim, Tae Geun Kim, Seong Chan Park, Stanislav Rusak, and Minoru Tanaka for useful comments.
The work of Y.H., H.K., and K.O.\ are supported in part by JSPS KAKENHI Grant Nos.\ 16J06151 (YH), 16K05322 (HK), and 15K05053 (KO), respectively.

\appendix
\section{Renormalization group equations}\label{calculation}

\begin{table}
\begin{center}
\begingroup
\renewcommand{\arraystretch}{1.2}
\begin{tabular}{l|lc}
						  & Value					   & Reference\\\hline
Planck mass	$\MP$ & $2.4353\times10^{18}$ GeV & \cite{Olive:2016xmw}\\
Higgs mass				  & 125.09 GeV				   & \cite{Olive:2016xmw}\\
$Z$ bozon mass $M_Z$	  & 91.1876 GeV				   & \cite{Olive:2016xmw}\\
$\alpha_3 (M_Z)$		  & 0.1184					   & \cite{Buttazzo:2013uya}\\
The expectation value of the
Higgs field $v$		  	  & 246 GeV					   & \cite{Olive:2016xmw}\\
\end{tabular}
\endgroup
\end{center}
\caption{Boundary condition for the RGEs.\label{table:phys}}
\end{table}

We have calculated the lower bound on $r$ as follows:
\begin{enumerate}
\item Solve the RGEs~\eqref{eq:rge-first}--\eqref{eq:rge-last} (shown in the end of this section) for given parameters.
The effects from right-handed neutrino is introduced only at high energy scale $\varphi\geq\MR$: We set $n_\nu=0$ and $\MR=0$ in $\varphi<\MR$.
As the boundary condition to solve the RGEs, we have used Eqs.~(2.11)--(2.15) in \cite{Hamada:2014xka} and the values in Table~\ref{table:phys}.\ef{\footnote{\ef{
This is based on Ref.~\cite{Buttazzo:2013uya}, where the top Yukawa coupling and the electroweak gauge couplings are extracted with full 2-loop NNLO precision.
}}}
\item Calculate the one-loop effective Higgs potential\al{
V_{\varphi\leq\Lambda}={\lambda_\eff\ov 4} \varphi^4
}where
\al{
\lambda_\eff
	&:=	e^{4\Gamma}\Bigg[\lambda
		+{1\over16\pi^2}\Bigg\{-3y_t^4\bigg(\ln{y_t^2\over2}-{3\over2}+2\Gamma\bigg)
		+{3g_2^2\over8}\bigg(\ln{g_2^2\over4}-{5\over6}+2\Gamma\bigg)\nn
	&\quad\phantom{e^{4\Gamma}\Bigg[}
		+{3(g_Y^2+g_2^2)^2\over16}\bigg(\ln{g_Y^2+g_2^2\over4}-{5\over6}+2\Gamma\bigg)
		-n_\nu y_\nu^4\ln\fn{\MR+\sqrt{\MR^2+4y_\nu^2 \varphi^2}\ov2\sqrt{\MR^2+\varphi^2}}
		\Bigg\}\Bigg]
\label{eq:lambda-eff}
}
is the effective Higgs-self coupling.\footnote{\label{roughness}
The last term in the braces is introduced to \emph{naively} take into account the effect of the neutrino loop on the effective potential. We have checked that its effect is at most few percent.
}
We set $\mu=\varphi$ when we calculate $\lambda_\eff$;
see \cite{Hamada:2014wna,Hamada:2016onh} for the details about the renormalization scale $\mu$ of $\lambda_\eff$.
The one-loop wave-function renormalization
\al{
\Gamma\fn{\varphi}
	&=	\int^\varphi_{m_t}{1\over16\pi^2}\paren{{9\over4}g_2^2+{3\over4}g_Y^2-3y_t^2-n_\nu y_\nu^2}\df\ln\mu
		\label{Gamma equation}
}
\ef{is taken into account.}
\item Change the parameters using the false position method until the value of the potential minimum becomes sufficiently close to zero.
\item Calculate the maximum value of potential and obtain $r_\tx{bound}$ via Eq.~\eqref{eq:rmin}.
\end{enumerate}
We have obtained the RGEs for arbitrary $n_\nu$ combining the $n_\nu=1$ RGEs~\cite{Haba:2014zja} and the $n_\nu=3$ ones~\cite{Kawana:2014zxa}:
\al{
\frac{\df g_Y}{\df\ln\mu}
	&=	{1\over16\pi^2}\frac{41}{6}g_Y^3+\frac{g_Y^3}{(16\pi^2)^2}\left({199\over18}g_Y^2+{9\over2}g_2^2+{44\over3}g_3^2-{17\over6}y_t^2-{n_\nu\over2}y_\nu^2\right),
\label{eq:rge-first}\\
\frac{\df g_2}{\df\ln\mu}
	&=	-{1\over16\pi^2}\frac{19}{6}g_2^3
		+\frac{g_2^3}{(16\pi^2)^2}\left({3\over2}g_Y^2+{35\over6}g_2^2+12g_3^2-{3\over2}y_t^2-{n_\nu\over2}y_\nu^2\right),\\
\frac{\df g_3}{\df\ln\mu}
	&=	-\frac{7}{16\pi^2}g_3^3+\frac{g_3^3}{(16\pi^2)^2}\left({11\over6}g_Y^2+{9\over2}g_2^2-26g_3^2-2y_t^2\right),\\
\frac{\df y_t}{\df\ln\mu}
	&=	\frac{y_t}{16\pi^2}\bigg(\frac{9}{2}y_t^2+n_\nu y_\nu^2-\frac{17}{12}g_Y^2-\frac{9}{4}g_2^2-8g_3^2\bigg)+\frac{y_t}{(16\pi^2)^2}\bigg\{-12y_t^2-{9n_\nu\over4}y_\nu^4-{9n_\nu\over4}y_t^2 y_\nu^2\nn
	&\quad
		+6\lambda^2+{\frac{1}{4}\kappa^2}-12\lambda y_t^2+g_Y^2\paren{\frac{131}{16}y_t^2+{5n_\nu\over8}y_\nu^2}+g_2^2\paren{\frac{225}{16}y_t^2+{5n_\nu\over8}y_\nu^2}+36 g_3^2 y_t^2\nn
	&\quad
		+\frac{1187}{216}g_Y^4-\frac{23}{4}g_2^4-108g_3^4-\frac{3}{4}g_Y^2 g_2^2+9g_2^2 g_3^2+\frac{19}{9}g_3^2 g_Y^2\bigg\},\\
\frac{\df \lambda}{\df\ln\mu}
	&=	\frac{1}{16\pi^2}\bigg(\frac{1}{2}\kappa^2+24 \lambda^2-3g_Y^2 \lambda-9g_2^2 \lambda+4n_\nu\lambda y_\nu^2+\frac{3}{8}g_Y^4+\frac{3}{4}g_Y^2 g_2^2 +\frac{9}{8}g_2^4+12\lambda y_t^2-6y_t^4-2n_\nu y_\nu^4\bigg)\nn
	&\quad
		+\frac{1}{(16\pi^2)^2}\bigg\{-2\kappa^3-5\kappa^2\lambda-312\lambda^3+36\lambda^2(g_Y^2+3g_2^2)
			-\lambda\left({629\over24}g_Y^4-{39\over4}g_Y^2g_2^2+{73\over8}g_2^4\right)\nn
	&\phantom{\quad+\frac{1}{(16\pi^2)^2}\bigg\{}
		+\frac{305}{16}g_2^6-\frac{289}{48}g_Y^2 g_2^4 -\frac{559}{48}g_Y^4 g_2^2 -\frac{379}{48}g_Y^6 -32 g_3^2 y_t^4-\frac{8}{3}g_Y^2 y_t^4-\frac{9}{4}g_2^4 y_t^2-{3n_\nu\over4}g_2^4y_\nu^2\nn
	&\phantom{\quad+\frac{1}{(16\pi^2)^2}\bigg\{}
		+\lambda y_t^2 \bigg(\frac{85}{6}g_Y^2+\frac{45}{2}g_2^2+80g_3^2\bigg)+\lambda y_\nu^2\paren{{5n_\nu\over2}g_Y^2+{15n_\nu\over2}g_2^2}+g_Y^2 y_t^2\bigg(-\frac{19}{4}g_Y^2+\frac{21}{2}g_2^2\bigg)\nn
	&\phantom{\quad+\frac{1}{(16\pi^2)^2}\bigg\{}
		-g_Y^2 y_\nu^2\paren{{n_\nu\over4}g_Y^2+{n_\nu\over2}g_2^2}-144 \lambda^2 y_t^2-48n_\nu y_\nu^2-3\lambda y_t^4-n_\nu\lambda y_\nu^4+30y_t^6+10n_\nu y_\nu^6\bigg\},
}
\al{
\frac{\df y_\nu}{\df\ln\mu}
	&=	\frac{y_\nu}{16\pi^2}\bigg\{\paren{n_\nu+{3\over2}}y_\nu^2+3y_t^2-{3\over4}g_Y^2-{9\over4}g_2^2\bigg\}+\frac{y_\nu}{(16\pi^2)^2}\bigg\{-\paren{{9n_\nu\over2}-{3\over2}}y_\nu^4-{27\over4}y_t^4\nn
	&\quad
		-{27\over4}y_t^2 y_\nu^2+6\lambda^2+{1\over4}\kappa^2
		-12\lambda y_\nu^2
		+g_Y^2\paren{\paren{{5n_\nu\over8}+{93\over16}}y_\nu^2+{85\over24}y_t^2}\nn
	&\quad
		+g_2^2\paren{\paren{{15n_\nu\over8}+{135\over16}}y_\nu^2+{45\over8}y_t^2}+20g_3^2 y_t^2+{35\over24}g_Y^4-{23\over4}g_2^4-{9\over4}g_Y^2 g_2^2\bigg\},\\
\frac{\df \kappa}{\df\ln\mu}	
	&=
        \frac{\kappa}{16\pi^2}\left(12\lambda+ \lambda_S+4\kappa+6y_t^2+2n_\nu y_\nu^2-\frac{3}{2}g_Y^2 -\frac{9}{2}g_2^2  \right)
        +\frac{\kappa}{(16\pi^2)^2}\bigg\{-\frac{21}{2}\kappa^2\nn
	&\quad
		-72\kappa\lambda-60\lambda^2
		-6\kappa\lambda_S-\frac{5}{6}\lambda_S^2-y_t^2(12\kappa +72\lambda)-4n_\nu\kappa y_\nu^2-24n_\nu\lambda y_\nu^2\nn
	&\quad
		-\frac{27}{2} y_t^4-{9n_\nu\over2}y_\nu^4+g_Y^2 (\kappa +24\lambda)
		+ g_2^2(3\kappa+72\lambda)+y_t^2\left(\frac{85}{12} g_Y^2+\frac{45}{4} g_2^2+40 g_3^2\right)\nn
	&\quad
		+y_\nu^2\paren{{5n_\nu\over4}g_Y^2+{15n_\nu\over4}g_2^2}
		+{557\over48}g_Y^4-{145\over16}g_2^4+\frac{15}{8}g_Y^2 g_2^2\bigg\},\\
\frac{\df \lambda_S}{\df\ln\mu}	
        &=\frac{1}{16\pi^2}\left(3\lambda_S^2+12 \kappa^2\right)
        +\frac{1}{(16\pi^2)^2}\bigg\{
			-\frac{17}{3}\lambda_S^3\nn
        &\quad
        	-20\kappa^2 \lambda_S-48\kappa^3
			-72\kappa^2 y_t^2 -24n_\nu\kappa^2 y_\nu^2
			+24\kappa^2 g_Y^2 +72\kappa^2 g_2^2\bigg\}.
\label{eq:rge-last}
}


\section{Explaining the form of envelope by potential shape}\label{OHP}

In this section, we explain the shape of envelopes denoted by the black or colored-thick line in the figures.

\begin{figure}[b]
  \begin{center}
   \includegraphics[height=10em]{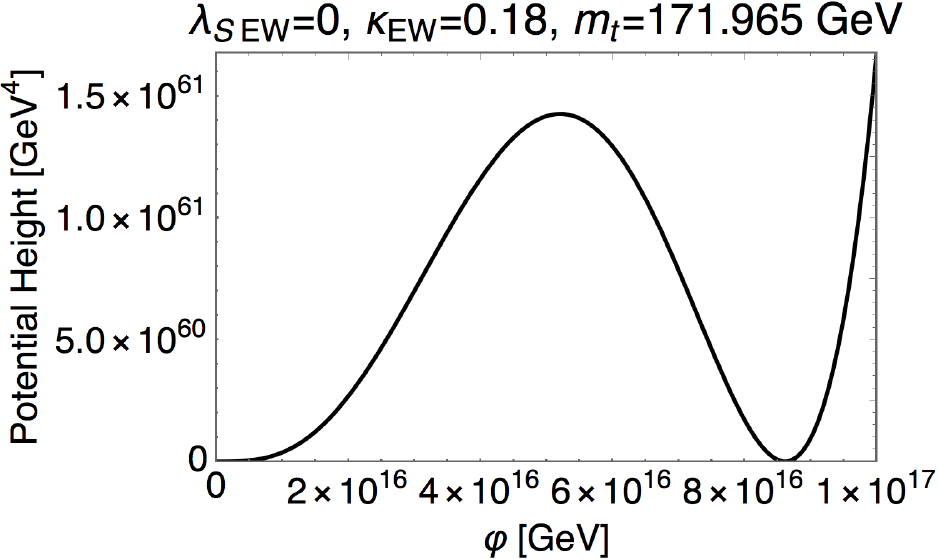}
  \end{center}
  \caption{The shape of Higgs potential with the values of $m_t$ and $m_\tx{DM}$ ($\kappa$) that corresponds to a point near the minimum of the envelope in the left of Fig.~\ref{fig:N0}.}
  \label{fig:N0-min}
\end{figure}

Let us start with the case without heavy right-handed neutrino (the left of Fig.~\ref{fig:N0}).
At the minimum point of envelope, the maximum value of the Higgs potential becomes smallest.
In Fig.~\ref{fig:N0-min}, we show the shape of potential near this point.
At this minimum point, the height at the local maximum (at $\varphi\simeq 5\times10^{16}\GeV$ in the case of Fig.~\ref{fig:N0-min}) becomes identical to the height at $\varphi=\Lambda$.
If the local potential minimum (at $\varphi=9\times10^{16}\GeV$ in the case of Fig.~\ref{fig:N0-min}) is moved left, then the height at $\varphi=\Lambda$ becomes larger, while moved right, the height at the local maximum becomes larger.
The left of the minimum point of the envelope is governed by the local maximum of the potential, while the right by the value at $\varphi=\Lambda$.

\begin{figure}[t]
  \begin{center}
   \includegraphics[height=7.9em]{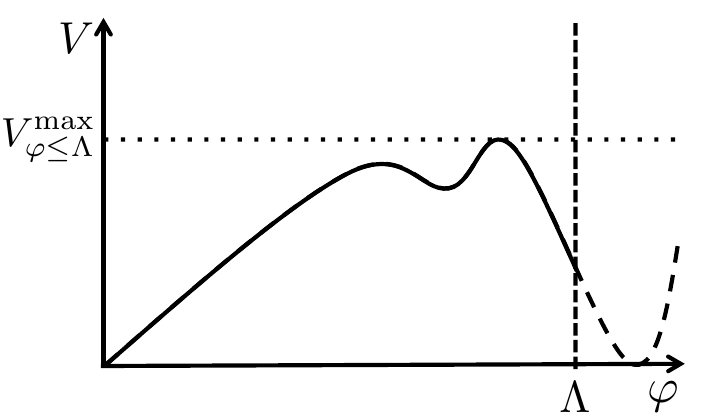}\hfill
   \includegraphics[height=7.9em]{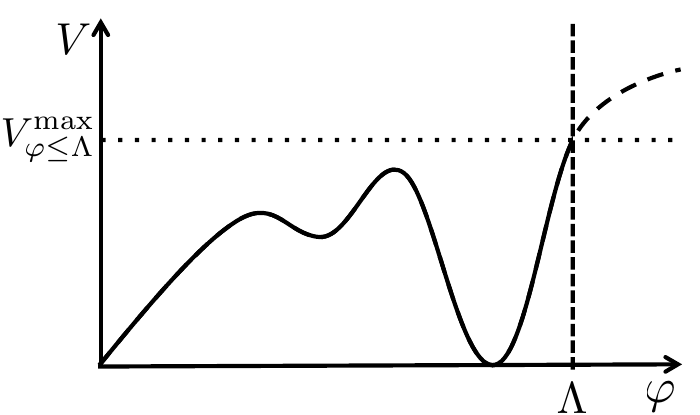}\hfill
   \includegraphics[height=7.9em]{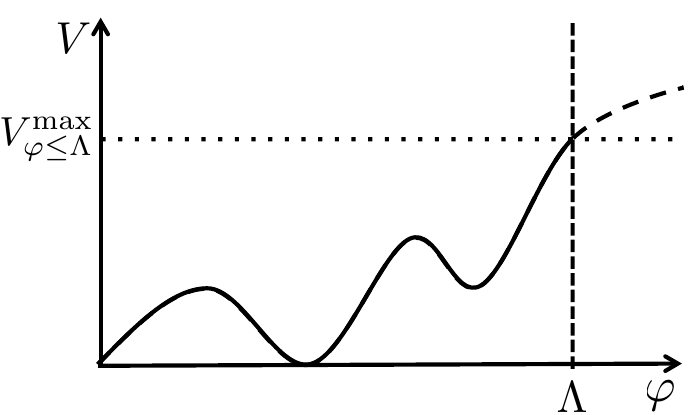}
  \end{center}
 \caption{Left: Typical potential shape when its minimum is at $\varphi>\Lambda$.
 Center: Typical potential shape when the minimum at larger $\varphi$ gives lower height.
 Right: Typical potential shape when the minimum at smaller $\varphi$ gives lower height.\label{fig:over17}}
\end{figure}

\magenta{
Now we turn to the case with right-handed neutrinos.
The envelope is denoted by the colored thick line in Figs.~\ref{fig:N1}--\ref{fig:N3}.
We note that each envelope in Fig.~\ref{fig:houraku4sigma} has, in addition to the cusp, one more (hardly seeable) non-smooth point, which is located, e.g.\ in the $n_\nu=1$ panel, at
$\paren{m_\tx{DM},\log_{10}r}\sim\paren{1050\GeV,-1.5}$ for $\MR=10^{14}\GeV$, 
$\paren{1100\GeV,-1.7}$ for $\MR=10^{14.1}\GeV$,
$\paren{1150\GeV,-2}$ for $\MR=10^{14.2}\GeV$, and
$\paren{1220\GeV,-2.6}$ for $\MR=10^{14.3}\GeV$.
This is because the Higgs potential has two local minima in general: The one at higher (lower) $\varphi$ is due to the neutrino (top quark) contribution.\footnote{
Here we let the word ``minimum'' also stands for mere a concavity, namely, even when it is not really a minimum.
}
There are the following three kinds of potential shapes:
\begin{enumerate}[(i)]
\item \label{higher vacuum goes beyond} On the left side of the cusp of each envelope, the potential minimum at higher $\varphi$ is located at $\varphi>\Lambda$; see the left of Fig.~\ref{fig:over17}.
\item \label{New Hamada Paradise} In between the cusp and the non-smooth point of each envelope, there are two potential minima and the height of the one at larger $\varphi$ is smaller; see the center of Fig.~\ref{fig:over17}.
\item \label{Old Hamada Paradise}
On the right side of the  non-smooth point of each envelope, there are two potential minima and the height of the one at lower $\varphi$ is smaller; see the right of Fig.~\ref{fig:over17}.
\end{enumerate}
These two potential minima are degenerate on the black line in Figs.~\ref{fig:houraku2sigma} and \ref{fig:houraku4sigma}, and the cases~\eqref{New Hamada Paradise} and \eqref{Old Hamada Paradise} become identical.

\bibliographystyle{TitleAndArxiv}
\bibliography{refs}

\end{document}